\documentclass[lettersize,journal]{IEEEtran}
\usepackage{amsmath,amsfonts}
\usepackage{algorithmic}
\usepackage{algorithm}
\usepackage{array}
\usepackage[caption=false,font=normalsize,labelfont=sf,textfont=sf]{subfig}
\usepackage{textcomp}
\usepackage{stfloats}
\usepackage{bibentry}
\usepackage{url}
\usepackage{verbatim}
\usepackage{graphicx}
\usepackage{cite}
\usepackage{multirow}
\usepackage{balance}
\usepackage{subfloat}
\usepackage{threeparttable}
\usepackage{makecell}
\usepackage{stfloats}
\usepackage{amssymb}    
\usepackage{bbding}     
\usepackage[utf8]{inputenc}
\usepackage{amsmath}
\usepackage{pifont}
\usepackage{booktabs}
\usepackage{array}
\usepackage{siunitx}
\usepackage{xspace}
\usepackage{ragged2e}

\def\BibTeX{{\rm B\kern-.05em{\sc i\kern-.025em b}\kern-.08em
		T\kern-.1667em\lower.7ex\hbox{E}\kern-.125emX}}
\usepackage{balance}
\begin{document}
			\title{A Multi-Modal  Channel Sounding System  for Environment-Aware Channel Measurements in 6G Dynamic Scenarios}
	\author{Xuejian Zhang,~\IEEEmembership{Student Member,~IEEE,} Ruisi He,~\IEEEmembership{Senior Member,~IEEE,}   \\
		Mi Yang,~\IEEEmembership{Member,~IEEE,} 
		Zhengyu Zhang,~\IEEEmembership{Student Member,~IEEE,}  
			Ziyi Qi
	\thanks{

This work is supported by 
the Fundamental Research Funds for the Central Universities under Grant 2022JBQY004, 
the National Natural Science Foundation of China under Grant 62271037, 62431003, 62671034, and 62301022, 
and the Postdoctoral Fellowship Program (Grade C) of China Postdoctoral Science Foundation under Grant GZC20261923.
\emph{(Corresponding authors: Ruisi He; Zhengyu Zhang.)}

X. Zhang, R. He, M. Yang, Z. Zhang,  and Z. Qi are with the State Key Laboratory of Advanced Rail Autonomous Operation, the School of Electronics and Information Engineering, and the Frontiers Science Center for Smart High-speed Railway System, Beijing Jiaotong University, Beijing 100044, China
(email: 23115029@bjtu.edu.cn; ruisi.he@bjtu.edu.cn;  myang@bjtu.edu.cn; 21111040@bjtu.edu.cn; 22115006@bjtu.edu.cn).

}
}
	\maketitle

\begin{abstract}

The deep integration of communication and sensing is widely recognized as a core 6G vision, underscoring the importance of comprehensive environment awareness. 
Accurate channel modeling forms the foundation of 6G system design and optimization, and channel sounders provide the essential empirical basis.
However, existing channel sounders, although supporting wide bandwidth and large antenna arrays in selected bands, generally lack cross-band capability, struggle in dynamic scenarios, and provide limited environmental awareness. 
The absence of detailed environmental information restricts the development of environment-aware channel models.
To address this gap, we propose a multi-modal sensing and channel sounding fusion system that enables temporally and spatially synchronized acquisition of  images, point clouds, geolocation information, and multi-band multi-antenna channel data.
The modular architecture facilitates rapid deployment in diverse dynamic environments.
The system supports 10 MHz--44 GHz  operation with up to  1 GHz  bandwidth and 1 ns  delay resolution, enabling multi-antenna measurements with a minimum switching interval of 8 $ \mu s$. 
Moreover, it provides centimeter-level 360$^{\circ}$ environmental sensing and meter-level positioning accuracy. Key metrological performance metrics, including an approximately 130 dB  measurable propagation loss at  28 GHz, phase-coherent acquisition, calibrated channel response, LiDAR geometric correction, and bounded multi-modal synchronization margins, are validated through a vehicle-to-infrastructure measurement campaign. The established system provides a scalable experimental platform for environment-aware channel modeling, facilitating the development and evaluation of integrated sensing and communication technologies in future 6G dynamic scenarios.

\end{abstract}


\section{Introduction}
\IEEEPARstart{6}{G} 
is envisioned to realize the deep convergence of communication, sensing, and artificial intelligence (AI), thereby enabling intelligent, ultra-reliable, and low-latency services for key application scenarios such as vehicle-to-everything (V2X), high-speed railway, and unmanned aerial vehicle (UAV) communications \cite{he2024, zzy2023, lei2023}.
Unlike previous generations primarily focused on communication performance, 6G aims to equip wireless systems with environmental awareness and context-adaptive capabilities, a vision widely recognized in both academia and industry \cite{itu1, hbx, COST}.

Accurate  channel modeling underpins 6G system design, performance optimization, and algorithm validation \cite{zxj2025-2}. 
In emerging 6G scenarios featuring high mobility and complex environmental dynamics, such as V2X communications, wireless channels exhibit pronounced non-stationarity and rapid temporal variation, with channel evolution intrinsically coupled with the evolution of the physical propagation environment \cite{mao2024}.
However, traditional statistical channel modeling relies solely on  radio frequency (RF) measurement data, making it difficult to capture fine-grained environmental dynamics and to establish explicit mappings between environmental evolution and channel characteristics \cite{zxj2025}. 
As a result, existing channel models often suffer from limited physical interpretability, weak generalization capability, and low prediction accuracy, hindering their applicability to environment-aware intelligent communications in 6G \cite{gu2026}.

The rapid advancement of multi-modal sensing technologies offers a promising pathway to overcome the above limitations \cite{imran2024, feifei2023}. 
Heterogeneous sensors, including Light Detection and Ranging (LiDAR), panoramic cameras, and global navigation satellite system (GNSS) modules, can provide complementary and high-resolution descriptions of the physical propagation environment \cite{wang2026}. 
As these sensing data and wireless channel measurements originate from the same physical world, they exhibit strong spatio-temporal consistency and correlation, enabling a more comprehensive understanding of environmental dynamics \cite{bl2025}. 
However, a systematic experimental platform capable of accurate spatio-temporally synchronized acquisition, unified calibration, and integrated fusion of multi-modal sensing data and wideband channel measurements remains largely absent \cite{cx-2024, deepsense}. 
Existing systems are typically developed independently for either channel sounding or environmental sensing, resulting in fragmented interfaces, inconsistent timestamps, incompatible data formats, and limited synchronization accuracy, which severely hinder research on environment–channel joint modeling for 6G integrated sensing and communication (ISAC) systems \cite{CJE}.

\subsection{Related Work}

\begin{table*}
	\belowrulesep=0pt
	\aboverulesep=0pt
    \setlength\tabcolsep{4pt}    
	\renewcommand{\arraystretch}{1.2} 
	\begin{center}
		\caption{Comparison of Related Works on Environment Sensing and Channel Sounding Platform.}
		\label{review}
		\begin{threeparttable}
			\begin{tabular}{ c| c| c| c| c| c| c| c| c| c| c}
			\toprule
			\multirow{2}{*}{\textbf{\makecell[c]{Ref.}}}  & \multirow{2}{*}{\textbf{Scenarios}} & 
            \multicolumn{3}{c|}{\textbf{Environment Sensing Data}} &  \multicolumn{4}{c|}{\textbf{Channel Sounding Data}} & \multirow{2}{*}{\textbf{Mobility}} & 
            \multirow{2}{*}{\makecell[c]{\textbf{Sounder Type}}}  \\
			\cline{3-9}
			~   &  ~  &  \makecell[c]{Image} & \makecell[c]{ Point Cloud } &  \makecell[c]{ Geolocation }  &  Freq.  & Bandwidth & Antenna Setup  & \makecell[c]{ Waveform}&  ~ & ~   \\
			\midrule   
            \cite{sm2025}  & Indoor ISAC &  \ding{53} & \ding{53} & \ding{53} & \makecell[c]{5--6 GHz} & 400 MHz  &  128×16 MIMO & ZC & Low Speed & USRP \\
			\hline
            \cite{wang2023usrp}  & UMi/O2I &  \ding{53} & \ding{53} & \ding{53} & \makecell[c]{4--6 GHz} & 180 MHz  &  SISO & PN & Low Speed & USRP \\
			\hline
            \cite{bas2018real}  & Indoor &  \ding{53} & \ding{53} & \ding{53} & 3--18 GHz & 1 GHz  &  SISO &  Multi-Tone & 5 km/h & NI PXIe \\
			\hline
            \cite{cxs2024}  & Indoor &  \ding{53} & \ding{53} & \ding{53} & \makecell[c]{28 GHz} & 2 GHz  & 128×256 MIMO & ZC & 0.36 km/h & NI PXIe \\
			\hline
            \cite{bas2019}  & UMi &  \ding{53} & \ding{53} & \ding{53} & 28 GHz & 400 MHz  &  16×16 MIMO &  Multi-Tone & \ding{53} & NI PXIe \\
			\hline
            \cite{cai2020dynamic}  & Indoor &  \ding{53} & \ding{53} & \ding{53} & \makecell[c]{28--30 GHz} & 200 MHz  & 1×360 SIMO & CW & \ding{53} & VNA \\
			\hline
            \cite{al2024}  & Indoor ISAC &  \ding{53} & \ding{53} & \ding{53} & 28 GHz & 2 GHz  &  Rotating-Mirror &  ZC & \ding{53} & FPGA Board \\
            \hline
             \cite{huang2020multi}  &  V2V &  \ding{53} & \ding{53} & \checkmark & 26.5--40 GHz & 320 MHz  &  4×4 MIMO &  PN & 60 km/h & Keysight \\
			\hline
            \cite{mao2023uav}  & A2G &  \ding{53} & \ding{53} & \checkmark & \makecell[c]{2.4/3.5 GHz} & 100 MHz  &  1×8 SIMO &  ZC & 7.2 km/h & NI PXIe \\
			\hline
            \cite{chen2023passive}  & \makecell[c]{ V2I } &   \ding{53} & \ding{53} & \checkmark & 3.5 GHz & 100 MHz  &SISO & 5G & 40 km/h & USRP \\
            \hline
             \cite{wu2021measurement}  & \makecell[c]{ UMi } &   \ding{53} & \ding{53} & \checkmark & 3.5--3.6 GHz & 122.88 MHz  & 1×16 SIMO & 5G & Low Speed & USRP \\
            \hline
            \cite{miao2023sub}  & UMi/O2I &  \ding{53} & \ding{53} & \checkmark & \makecell[c]{3--40 GHz} & 200/1000 MHz  &  SISO & PN & Low Speed & R\&S \\
			\hline
            \cite{kim2022}  & UMa/UMi &  \checkmark & \ding{53} & \ding{53} & \makecell[c]{24/60 GHz} & 200/400 MHz  & 4×4 SIMO & Multi-Tone & \ding{53} & FPGA Board \\
			\hline
            \cite{gen2024}  & \makecell[c]{Indoor ISAC} &   \checkmark & \checkmark & \checkmark & 28/60 GHz & 2 GHz  & 8×16 MIMO & PN & \ding{53} & FPGA Board \\
            \hline
            
			\bottomrule
			\end{tabular}
			\vspace{0.2em}
            \parbox{\linewidth}{\footnotesize
	        \textit{Notes}:  
            PN: Pseudo-Noise sequence,
			ZC: Zadoff–Chu sequence,
	        CW: Continuous Wave.
	        VNA: Vector Network Analyzer,
	        UMi: Urban Microcell,
	        UMa: Urban Macrocell,
	        O2I: Outdoor-to-Indoor,
	        V2I: Vehicle-to-Infrastructure,
	        V2V: Vehicle-to-Vehicle,
	        A2G: Air-to-Ground,
	        MIMO: Multiple-Input Multiple-Output,
	        SIMO: Single-Input Multiple-Output,
	        SISO: Single-Input Single-Output.}
		\end{threeparttable} 
	\end{center}
\end{table*}

Existing channel sounders can be divided into three categories according to their technical evolution and functional capabilities, and the core progress and limitations of each category are summarized as in Table \ref{review}.

The first category consists of traditional wideband channel sounders targeting static or low-mobility scenarios, typically implemented on hardware such as universal software radio peripheral (USRP), National Instruments PCI eXtensions for Instrumentation Express (NI PXIe), and field-programmable gate array (FPGA).
For example, the USRP-based sounders in \cite{sm2025} and \cite{wang2023usrp} support MIMO and SISO channel measurements over 5–6 GHz and 4–6 GHz with bandwidths of 400 MHz and 180 MHz, respectively. 
A 3–18 GHz sounder with 1 GHz bandwidth for low-mobility indoor scenarios is presented in  \cite{bas2018real}.
At millimeter-wave (mmWave) frequencies,
\cite{cxs2024} develops a 128×256 massive MIMO channel sounder at 28 GHz with up to 2 GHz bandwidth. 
Sounders in  \cite{bas2019, cai2020dynamic} also operate at 28 GHz in MIMO/SIMO configurations but offer only 400 MHz and 200 MHz bandwidth and lack mobility support,
while \cite{al2024} designs a rotating-mirror-based  mmWave sounder with 2 GHz bandwidth.
Despite their strong performance in wideband and MIMO channel measurements, existing platforms have three limitations: single-band operation, limited support for dynamic outdoor scenarios, and the lack of environmental sensing for environment–channel joint modeling in 6G ISAC.

The second focuses on channel sounders developed for outdoor dynamic environments, typically incorporating GNSS modules to provide geolocation and synchronization for mobile applications.
For instance, \cite{huang2020multi} presents a 4×4 MIMO sounder for  V2V scenarios covering 26.5–40 GHz with mobility up to 60 km/h, while \cite{mao2023uav} introduces a 2.4/3.5 GHz   sounder for  A2G scenarios.
Moreover, passive sounders capable of receiving commercial 5G downlink signals are developed in \cite{chen2023passive} and \cite{wu2021measurement}. 
The wideband sounder in \cite{miao2023sub} further extends the operating range to 3–40 GHz with up to 1 GHz bandwidth.
Despite these advances, existing systems remain limited due to their reliance on coarse geolocation information and the absence of synchronized acquisition of environmental sensing and channel measurement data.

The third explores preliminary attempts to integrate environmental sensing into channel measurement systems to enrich context information. 
For example,
\cite{kim2022} integrates a panoramic camera into a 24/60 GHz double-directional sounder to capture environmental images. 
Similarly, \cite{gen2024} develops a context-aware sounder combining images, point clouds, and geolocation data for indoor ISAC scenarios.
Nevertheless, these systems remain constrained by single-band operation and limited mobility support in dynamic environments.

In summary, existing measurement systems still exhibit three key limitations that hinder environment-aware channel modeling for 6G. 
First, spectrum adaptability remains constrained, since most existing systems are band-specific and lack scalable wideband measurement capability across low-, mid-, and high-frequency ranges.
Second, support for dynamic scenarios remains inadequate, with many systems limited in mobility and unable to ensure stable data acquisition under rapidly time-varying propagation conditions. Third, environmental sensing is seldom integrated with channel sounding at the instrumentation level, leaving most systems unable to synchronously capture multi-modal environmental context and radio-channel measurements.

\subsection{Contributions}
To bridge these gaps, this paper proposes a multi-modal sensing and channel sounding system for environment-aware wireless measurements in 6G dynamic scenarios. The main contributions are summarized as follows.

\begin{itemize}
	\item[$\bullet$]  
     A cross-band wideband channel sounding architecture is developed, covering an ultra-wide frequency range from 10 MHz to 44 GHz. By integrating complementary VSG/VSA- and FlexRIO-based architectures with reconfigurable RF front ends, the proposed system overcomes the band-specific limitation of existing sounders and provides a scalable measurement platform for 6G full-spectrum heterogeneous networks.
	\item[$\bullet$]  A spatio-temporally synchronized multi-modal acquisition framework is established to jointly collect wireless channel responses, LiDAR point clouds, panoramic images, and geolocation information. Through rigid sensor mounting, spatial registration, timestamp-based alignment, and rubidium-clock timing reference, the system bridges electromagnetic channel measurement and physical environment sensing, enabling environment-aware channel characterization beyond standalone sounding or sensing platforms.

    \item[$\bullet$]  
    A modular and scalable hardware--software platform is implemented and quantitatively validated through calibration, system verification, and a representative dynamic  measurement campaign. The platform supports reconfigurable sensing modalities, frequency bands, and antenna arrays, and provides synchronized multi-modal datasets for dominant-scatterer interpretation, environment-aware channel modeling, ISAC studies, and AI-assisted channel prediction in 6G dynamic scenarios.
   
\end{itemize}

The remainder of this paper is organized as follows. Section II presents the architecture of the multi-modal fusion sensing system. Sections III and IV detail the channel sounding and visual sensing subsystems, respectively. Section V evaluates the performance of the proposed system, and Section VI presents a representative measurement campaign. Finally, Section VII draws the conclusions.

\section{Architecture of Multi-Modal  Sensing and Channel Sounding  System}  
This section presents the architecture, design principles, and key specifications of the proposed system. The architecture is conceived as a fully integrated measurement instrument that jointly acquires wideband channel responses and multi-modal environmental information, rather than as a simple combination of independent sensing and sounding platforms. Through a unified spatio-temporal synchronization framework, each channel snapshot is aligned with its corresponding environmental observation, extending conventional standalone channel sounders toward environment-aware channel measurement.

\subsection{Architecture of System}
\begin{figure}[!t]
	\centering
	\includegraphics[width=.5\textwidth]{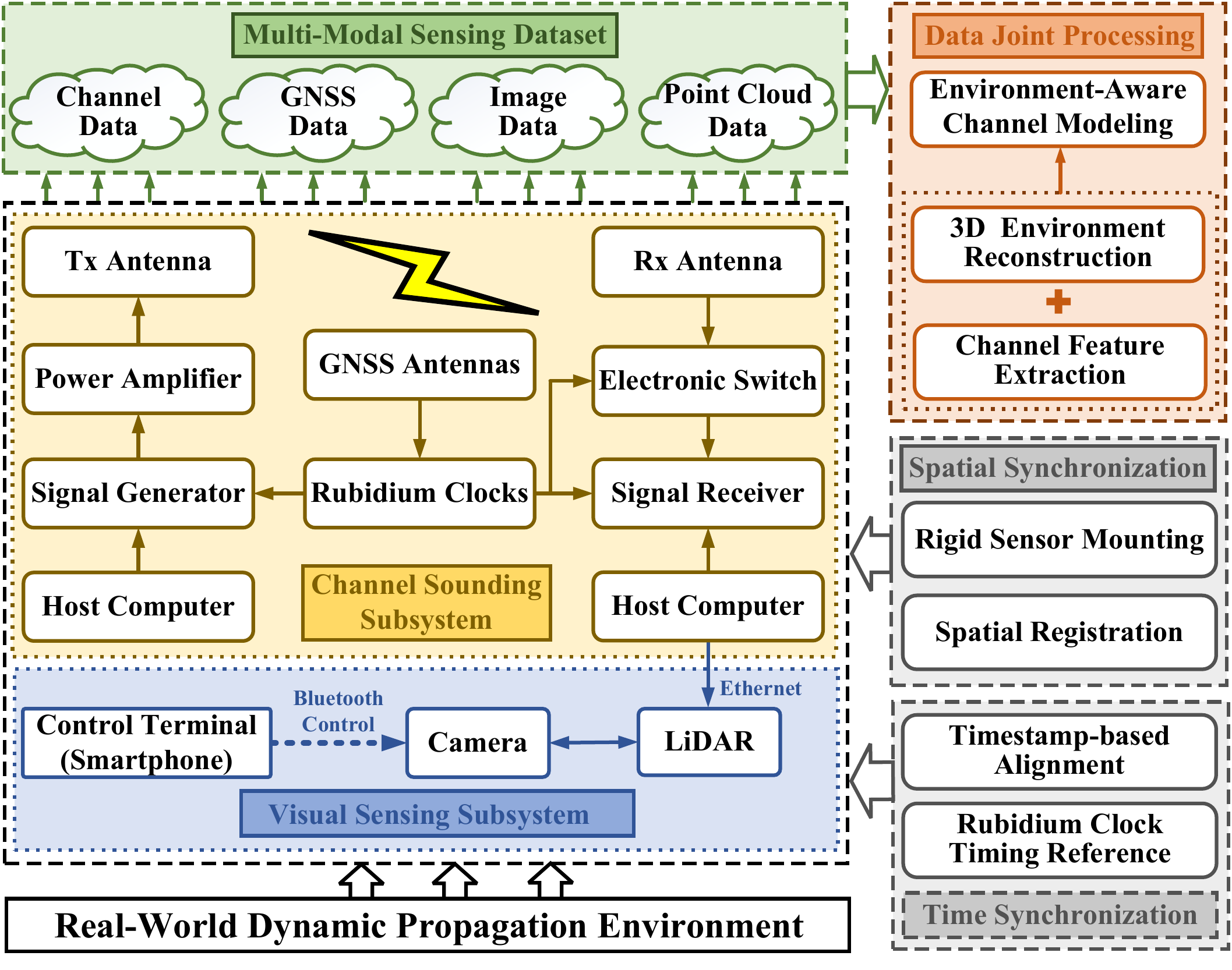}%
	\caption{Architecture of multi-modal channel sounding system.
	}                    
	\label{framework}
\end{figure}

Fig. \ref{framework} illustrates the overall architecture of the proposed  system, which comprises   channel sounding subsystem and   visual sensing subsystem. The former acquires wireless channel responses after electromagnetic propagation, whereas the latter captures direct representations of the physical environment. 

\begin{figure}[!t]
	\centering
	\includegraphics[width=.45\textwidth]{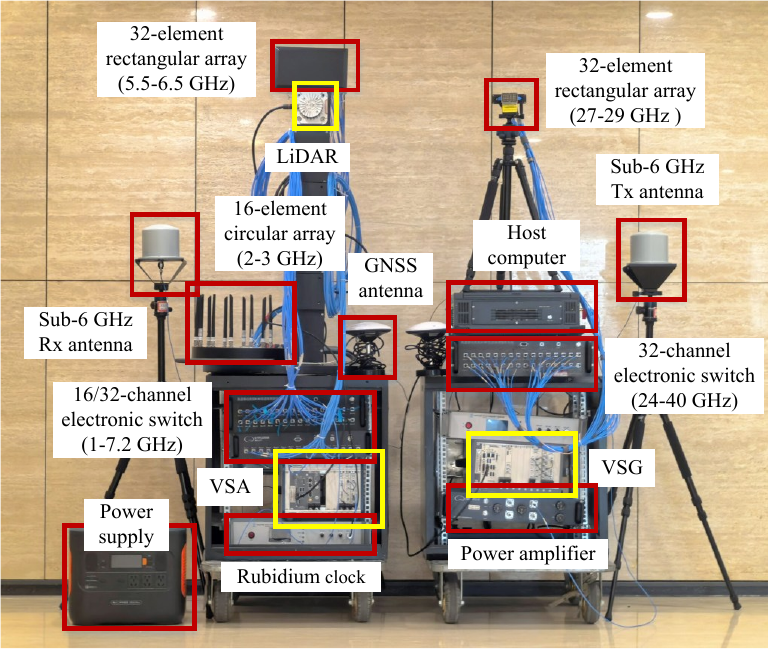}%
	\caption{Integrated hardware of the established platform showing the currently deployed configuration for Sub-6 GHz and 28 GHz mmWave measurements. 
	The system supports full 10 MHz--44 GHz operation through interchangeable modular components.
	}                    
	\label{device1}
\end{figure}

As shown in Fig. \ref{device1}, the integrated hardware platform adopts two NI-based channel sounding architectures to support multi-band measurements with limited hardware complexity: a vector signal generator/analyzer (VSG/VSA) architecture for Sub-6 GHz bands and a FlexRIO-based intermediate-frequency direct-sampling architecture for higher-frequency bands. 
Multiple antenna arrays enable multi-antenna spatio-temporal-frequency channel measurements, while Global Navigation Satellite System (GNSS) antennas synchronized with rubidium clocks provide geolocation information.
More details of this subsystem are presented in Section III.
Visual sensing subsystem integrates panoramic camera and LiDAR to acquire image and point cloud data of the same environment, which are spatially registered via a real-time fusion algorithm.
The LiDAR is interfaced with the receiver host computer via Ethernet for data acquisition and timestamp management, while the camera is remotely controlled through a Bluetooth-connected smartphone. The smartphone is used solely as an auxiliary camera-control terminal and, when available, provides optional GNSS metadata for visual data annotation. It does not serve as a timing reference, synchronization source, or independent sensing module.
The detailed design of this subsystem is described in Section IV.

Given the use of heterogeneous devices, spatio-temporal consistency across modalities is essential for reliable environment-aware channel analysis. As shown in Fig. \ref{framework}, four  mechanisms are adopted: rigid sensor mounting, spatial registration, timestamp-based alignment, and rubidium-clock timing reference. Specifically, rigid sensor mounting maintains fixed relative positions among the antennas, LiDAR, camera, and GNSS antenna; spatial registration transforms heterogeneous sensing data into a unified coordinate system; timestamp-based alignment associates channel snapshots, LiDAR frames, and camera images according to their temporal metadata; and the rubidium-clock timing reference provides a stable frequency and timing basis for coherent channel acquisition. After synchronization and alignment, the collected channel data, geolocation data, images, and point clouds form a unified multi-modal dataset that jointly characterizes the electromagnetic and physical environments. The dataset is then used for multipath parameter extraction, 3D environment reconstruction, and environment-aware channel modeling, completing an end-to-end workflow from data acquisition to fusion and modeling in dynamic scenarios.
Further details on the synchronization mechanism and data post-processing are provided in Section II-B and Section VI, respectively.

\subsection{Synchronization of Multi-Modal Data}
Ensuring the usability of multi-modal data requires temporal synchronization for proper alignment across devices and spatial synchronization for mapping into a unified 3D coordinate system \cite{Charan2024}.
Temporal synchronization is established through a hierarchical timing framework. 
For the channel sounding subsystem, two GNSS-disciplined rubidium clocks deployed at the transmitter (Tx) and receiver (Rx) provide stable 10 MHz reference signals and 
pulse per second (PPS) timing signals for coherent channel acquisition. At each side, the PXIe-6674T distributes the reference clock and trigger signals to the PXIe modules, ensuring consistent timestamping of channel snapshots. The geolocation data are also recorded with GNSS-referenced timestamps, providing a unified temporal basis for subsequent data alignment.

For the visual sensing subsystem, the LiDAR is connected to the host computer via Ethernet and records point clouds together with timestamp, GNSS, and  inertial measurement unit (IMU) information. 
The panoramic camera is operated through a Bluetooth-connected smartphone, which serves only as an auxiliary control terminal for starting and stopping video acquisition and, when available, for injecting GNSS metadata into the recorded visual data. It should be emphasized that neither Bluetooth nor Internet access is used as a precision clock-synchronization mechanism in the proposed system.

Spatial synchronization is achieved through a rigid mechanical integration design. As shown in Fig. 3, a circular magnetic mount with a unified interface and dedicated extension brackets is used to co-locate heterogeneous devices. The LiDAR and panoramic camera are mounted on an elevated bracket to avoid occlusion and ensure unobstructed LiDAR scanning and full $360^{\circ}$ image acquisition. 
The antennas are installed on a separate front bracket with a lower height to avoid blocking either line-of-sight (LoS) propagation or environmental sensing. 
The GNSS antenna is placed behind the LiDAR and aligned with the Rx antennas in the same horizontal plane. 
This rigid configuration provides a fixed spatial relationship among the channel sounder, LiDAR, camera, and GNSS modules, while also allowing additional sensing devices to be installed through the same mechanical interface.

At the data-processing level, multi-modal alignment is achieved through timestamp-based post-processing. Since different modalities operate at different sampling rates, channel snapshots, LiDAR frames, and camera frames are aligned according to their timestamps, frame indices, sampling intervals, and fixed spatial offsets.
Residual temporal mismatch is compensated through nearest-frame matching or interpolation.
For a typical urban V2I scenario with a vehicle speed of 40--60 km/h, a 50 ms frame-level mismatch corresponds to a spatial displacement of approximately 0.56--0.83 m, while even a conservative 100 ms bound results in only 1.11--1.67 m. 
This synchronization margin is sufficiently small for associating dominant channel clusters with large-scale environmental features such as walls, buildings, and vehicles, thereby supporting environment-aware channel analysis in dynamic scenarios.

  \begin{figure}[!t]
 	\centering
 	\includegraphics[width=.45\textwidth]{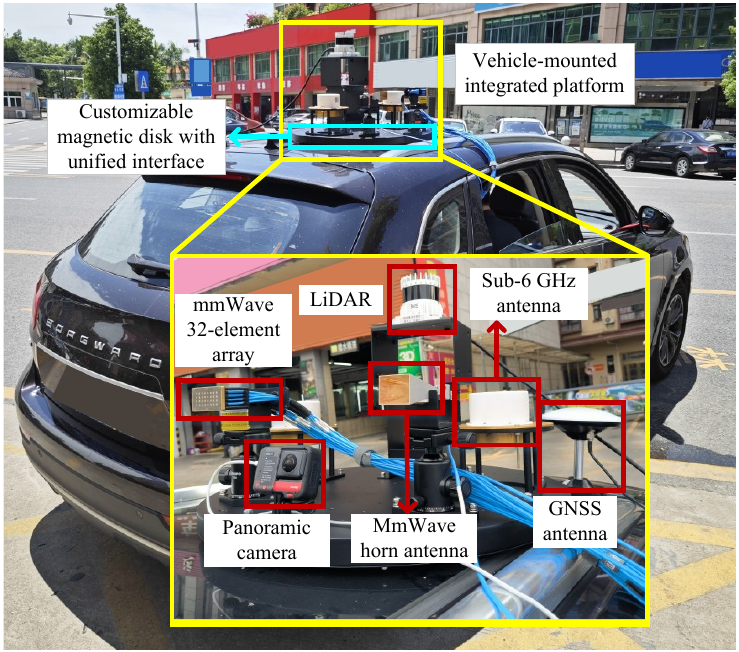}%
 	\caption{An example of spatial synchronization between multiple sensing devices using a magnetic disk in V2X scenarios.
 	}                    
 	\label{device5}
 \end{figure}

\subsection{System Specifications}
The proposed system  covers 10 MHz--44 GHz bands, with maximum transmit powers of 65 dBm  and bandwidths up to 1 GHz. 
It supports up to 100 channel snapshots/s with a minimum electronic switching interval of 8 $\mu s$ for SIMO measurements, ensuring fine temporal resolution. 
For environmental sensing, the LiDAR achieves a 200 m range with ±5 cm accuracy at 10 frames/s, while the camera captures 4K photos at up to 100 frames/s. 
The system achieves centimeter-level environmental sensing accuracy and meter-level positioning accuracy with a maximum position update rate of 120 Hz.
Detailed  specifications are listed in Table \ref{performance}, and the design and validations of each subsystem are presented   in Sections \uppercase\expandafter{\romannumeral3}, \uppercase\expandafter{\romannumeral4}, and \uppercase\expandafter{\romannumeral5}, respectively.


\begin{table}[!t]
\centering
\caption{Key Specifications of the Proposed Platform.}
\setlength{\tabcolsep}{7pt} 
\renewcommand{\arraystretch}{1.2}
\begin{tabular}{  c| c| c}
\hline
\multicolumn{3}{c}{\textbf{Radio Frequency Performance Metrics}} \\
\hline
 Frequency Bands & 10 MHz--6 GHz & 6 GHz--44 GHz\\
\hline
 Bandwidth & 50 MHz  & 1 GHz  \\ 
\hline
 Delay Resolution & 20 ns &  1 ns  \\
\hline
 Dynamic Range & 125 dB & 130 dB\\
\hline
 Antenna Array & \multicolumn{2}{c}{16/32-element } \\
\hline
 Transmit Power  &  Max. 47 dBm  & Max. 65 dBm  \\
\hline
 Switching Interval   &  \multicolumn{2}{c}{Min. 8 $\mu s$ }\\
 \hline
Acquisition Rate & 100 snapshots/s  & 50 snapshots/s \\
 \hline
 \hline
\multicolumn{3}{c}{\textbf{Environment-Aware Sensing Metrics}} \\

\hline
 Number of Beams     &   \multicolumn{2}{c}{128}  \\
\hline
Scanning Accuracy & \multicolumn{2}{c}{± 5 cm} \\
\hline
Scanning Range & \multicolumn{2}{c}{200 m} \\
\hline
Field of View & \multicolumn{2}{c}{360° (H) × 45° (V)} \\
\hline
Image Resolution & \multicolumn{2}{c}{Max. 4K} \\
\hline
Point Cloud Frame Rate & \multicolumn{2}{c}{10 Hz}  \\
\hline
Image Frame Rate & \multicolumn{2}{c}{Max. 100 Hz } \\
\hline
Additional Parameters   & \multicolumn{2}{c}{velocity, altitude, acceleration}\\
\hline
\hline
\multicolumn{3}{c}{\textbf{Geolocation and Synchronization   Metrics}} \\
\hline
Positioning Bands   & \multicolumn{2}{c}{  L1 (1575.42 MHz), B1 (1561.098 MHz) }  \\
\hline
Positioning Accuracy & \multicolumn{2}{c}{± 3 m} \\
\hline
Position Update Rate  &  \multicolumn{2}{c}{Max. 120 Hz}  \\
\hline
Timing Stability    &  \multicolumn{2}{c}{$\leq$ $1\times10^{-11}$}  \\
\hline
Supported Mobility    &  \multicolumn{2}{c}{$ \geq$ 100 km/h} \\
\hline
\end{tabular}
\label{performance}
\end{table}

%
%
%

%

\section{Channel Sounding Subsystem Design and Implementation} 
This section details the hardware design, signal processing workflow, and metrological performance of the channel sounding subsystem, which is responsible for the measurement of the first core measurand of the system: full-dimensional wireless channel characteristics. The measurement workflow of the subsystem is fully calibrated and verified to ensure the accuracy and repeatability of the channel measurement results, which meets the metrological requirements of wireless channel measurement instruments.

\subsection{Overall Architecture}
The channel sounding subsystem is primarily designed to generate, transmit, and receive vector sounding signals across multiple frequency bands for electromagnetic environment sensing. 
The overall architecture of the subsystem is illustrated in Fig. \ref{framework}. 
Its core components include a wideband signal generator and receiver. 
At the transmitter (Tx) side, the generated wideband signal is amplified by a power amplifier (PA) and radiated into wireless channel through  Tx antenna. 
At Rx side, the signal is captured and amplified by a low-noise amplifier (LNA), and optionally switched across multiple channels via high-speed electronic switches to enable time-division multiplexed multi-antenna measurements. 
The received signal is then recorded as in-phase and quadrature (IQ) data for offline processing. 
Time synchronization and positioning are ensured by  rubidium clocks disciplined by  GNSS reference.

To enable cross-band and large-bandwidth channel sounding, two complementary architectures are adopted. The Sub-6 GHz subsystem is implemented using the VSG/VSA architecture, whereas the 6–44 GHz measurements are realized through a FlexRIO-based intermediate-frequency (IF) direct-sampling architecture integrated with frequency up/down converters. Universal core modules, including the PXIe-5745 signal generator and PXIe-5775 digitizer, form a reusable baseband/IF processing platform, while band-dependent functions are implemented through interchangeable RF front-end modules, such as converters, switches, power amplifiers, and antennas. This decoupled design enables frequency extension without redesigning the core signal processing, timing synchronization, or data acquisition framework.

It should be clarified that the 10 MHz--44 GHz frequency range denotes the architectural frequency capability of the proposed modular platform, rather than the specific bands presented in Fig. \ref{device1}. 
The current implementation focuses on   sub-6 GHz bands and 28 GHz mmWave bands, which are representative candidates for 6G V2X and ISAC applications. 
Owing to the reconfigurable RF front end, the system can be extended to Frequency Range 2 (FR2) band (28--44 GHz) and Frequency Range 3 (FR3) band (7--24 GHz) by replacing only frequency-specific front-end modules, while retaining the core signal processing, timing synchronization, and data acquisition subsystems. 
By integrating the two architectures, the proposed subsystem supports channel measurements over 10 MHz--44 GHz with a maximum bandwidth of 1 GHz and a delay resolution of 1 ns, providing an extensible platform for future 6G full-spectrum measurements.

\subsection{Sounding Waveform}
Channel sounding transmits a known waveform and processes the received signal to extract the channel transfer function and impulse response \cite{sm2025}. 
Extending this to multiple antenna pairs enables spatial and directional characterization \cite{fd2022}.
In this work, a multi-tone signal $m(t)$ is adopted as  baseband waveform, expressed as \cite{bas2019}:
\begin{equation}
	m(t) = \sum\limits_{n=-N}^N e^{j(2\pi n \Delta f t + \theta_n)},
\end{equation}
where $\Delta f$ is the tone spacing, $2N+1$ is the number of tones, $t$ is time, $n$ is the tone index, and $\theta_n$ is the phase of the $n$-th tone. Out-of-band subcarriers are zeroed to confine power within the effective bandwidth, and the frequency-domain signal is converted to the time domain via inverse fast Fourier transform (IFFT). 
The resulting waveform exhibits a flat spectral profile, ensuring equal signal-to-noise ratio (SNR) across all tones and enabling accurate, unbiased channel estimation.

\begin{figure*}[!t]
	\centering
	\includegraphics[width=.98\textwidth]{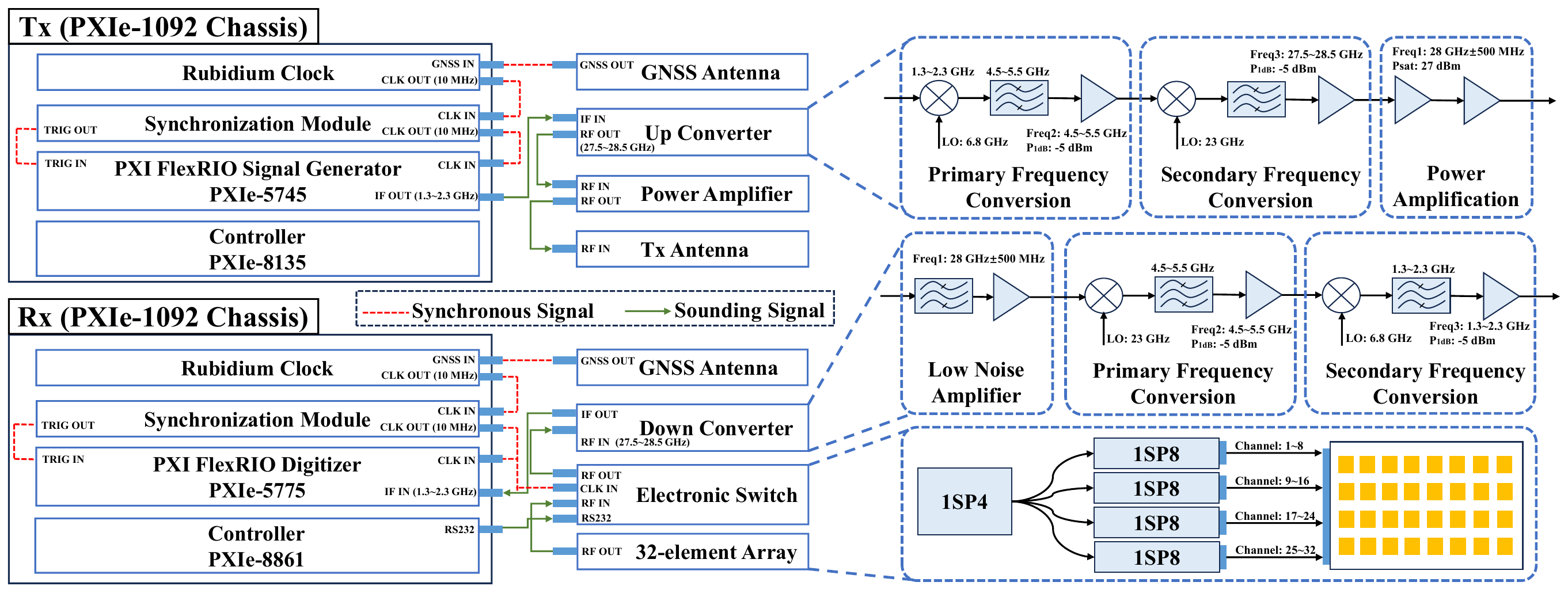}%
	\caption{Design and implementation  of mmWave channel sounder.
	}                    
	\label{block}
\end{figure*}

\subsection{Sounder Implementation}
The integrated channel sounding subsystem adopts a modular NI PXIe-based architecture, complemented by custom RF front-end components such as electronic switches, power amplifiers, up/down frequency converters, and rubidium clocks. This design satisfies the required measurement performance while preserving flexibility and reconfigurability across different frequency bands and scenarios.

Taking the 28 GHz mmWave sounder as an example, Fig. \ref{block} shows the system architecture. 
At Tx, a PXIe-5745 wideband signal generator produces the baseband probing waveform, which is then upconverted by a customized module. This module performs two-stage frequency conversion (1.3--2.3 GHz → 4.5--5.5 GHz → 27.5--28.5 GHz), followed by power amplification and radiation via a horn antenna.
On the receiver side, signals are collected by a 32-element antenna array through a high-speed 32-channel electronic switch and processed by a custom downconverter integrated with the switch in a 2U chassis. After filtering and low-noise amplification, the signal is downconverted back to 1.3--2.3 GHz and digitized by a PXIe-5775 wideband digitizer for storage and post-processing. 
Both upconversion and downconversion modules adopt a modular design, allowing reconfiguration by replacing mixers or power amplifiers to adjust frequency, bandwidth, or output power, thereby improving adaptability and reducing maintenance complexity.

\subsection{Sounder Hardware}

Fig. \ref{device1} presents a photograph of sounder hardware setup, and  key specifications of  major modules are summarized in Table \ref{hardware}. 
To avoid redundant parameter-level descriptions, this subsection focuses on the functional role and interconnection of each module.

\textit{1) VSA/VSG and FlexRIO modules.}
Two complementary PXIe-based architectures are adopted. 
The VSG/VSA-based architecture is used for Sub-6 GHz channel sounding, whereas the FlexRIO-based signal generator and digitizer support wideband IF direct sampling for the 6--44 GHz band when combined with external frequency converters. 
This dual-architecture design enables cross-band channel measurements while maintaining a   reconfigurable hardware implementation.

\begin{figure}[!t]
	\centering
	
	\includegraphics[width=.48\textwidth]{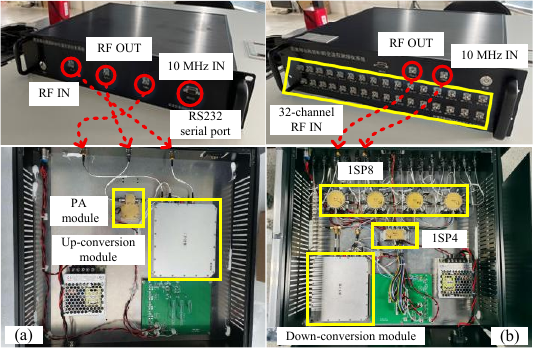}%
	
	\caption{
		(a) Up converter;
		(b) Down converter and electronic switch.
	}
	\label{converter}
\end{figure}

\textit{2) RF front-end.}
The RF front-end consists of power amplifiers, low-noise amplification, up/down converters, and electronic switches. 
The power amplifiers are selected according to the link-budget requirements of different frequency bands, while the up/down converters translate signals between the IF domain and  target RF bands. 
The modular converter design allows the operating frequency range, bandwidth, and output power to be reconfigured by replacing key RF components.
Two types of converters are developed. 
The first integrates a power amplifier and an electronic switch for the 27--29 GHz band,  as shown in Fig. \ref{converter}. 
The second is a standalone 6--44 GHz converter module used with an external power amplifier. 

\begin{table*}[!t]
	\belowrulesep=0pt
	\aboverulesep=0pt
	\renewcommand{\arraystretch}{1.5}
	\centering
	\caption{Key Hardware Components of Channel Sounding System.}
	\label{hardware}
	\begin{tabular}{c|c|c}
		\toprule
	     \textbf{Type} & \textbf{Model} & \textbf{Description} \\
		\midrule
		 VSG & NI PXIe-5673 &Freq: 50 MHz--6.6 GHz, Modulation bandwidth \textgreater 100 MHz\\
		\hline	
		 VSA & NI PXIe-5663 &Freq: 10 MHz--6.6 GHz, Max. BW: 50 MHz, IQ sampling rate: 150 MS/s \\
      \hline
		 \makecell[c]{FlexRIO \\ Signal Generator}       & NI PXIe-5745 &Freq: 1 MHz--3.2 GHz,  BW: 60 MHz--2.85 GHz, IQ sampling rate: 12-bit, 6.4 GS/s \\
		 \hline 
		 FlexRIO Digitizer    & NI PXIe-5775 & Freq: 1 MHz--6 GHz, BW: 500 kHz--6 GHz,  IQ sampling rate: 12-bit, 6.4 GS/s \\
		\hline
		 \makecell[c]{Synchronization \\ Module} & NI PXIe-6674T & Distributes a shared 10 MHz reference and trigger signals for precise Tx/Rx synchronization \\
		\hline
		 \multirow{2}{*}{Controller} & NI PXIe-8135 & 2.3 GHz dual-core CPU, Intel Core i7, 16 GB RAM, 512 GB HDD \\
		 \cline{2-3}
		 ~                    & NI PXIe-8861 & 2.8 GHz quad-core CPU, Intel Xeon Processor E3-1515M v5, 32 GB RAM, 1 TB HDD \\
		\hline
		 \multirow{2}{*}{PA} & HRTY-010060G-60W & Freq: 1 GHz--6 GHz, $P_{\text{sat}}$ = 47.5 dBm, $P_{\text{1dB}}$ = 45 dBm, 47.5 dB ± 5 dB gain\\
		 \cline{2-3}
		 ~ & LDPA27G36G30 & \makecell[c]{Freq: 27--36 GHz, $P_{\text{sat}}$ = 45 dBm, $P_{\text{1dB}}$ = 43 dBm, 45 dB ± 2 dB gain}   \\
		\hline
		 \multirow{3}{*}{Electronic Switch} & \multirow{3}{*}{Custom-designed} & Freq: 2--3 GHz, 16/32 channels, Min. switching interval: 8 $\mu s$,   Insertion loss:  4.5 dB  \\
		 \cline{3-3}
       ~ & ~ & Freq: 5.4--6.4 GHz, 32 channels,  LNA: 30 dB gain, Min. switching interval: 8 $\mu s$, Insertion loss:  8 dB  \\
       \cline{3-3}
		 ~ & ~ & \makecell[c]{Freq: 24 GHz--40 GHz, 32 channels,  Min. switching interval: 8 $\mu s$, Insertion loss:  10 dB} \\
		\hline
        Up-Converter &  Custom-designed  &  \makecell[c]{Type 1: two conversions: 1.3--2.3 GHz $\rightarrow$ 4.5--5.5 GHz $\rightarrow$ 27.5--28.5 GHz; \\
        Type 2: one conversion: 3.1--4.2 GHz $\rightarrow$ 6--44 GHz}\\
		\hline	
        Down-Converter & Custom-designed  & \makecell[c]{Two conversions: 27.5--28.5 GHz $\rightarrow$ 4.5--5.5 GHz $\rightarrow$ 1.3--2.3 GHz; \\
        Type 2: one conversion: 6--44 GHz $\rightarrow$ 3.1--4.2 GHz}\\
		\hline	
		 Rubidium Clock & SYN3204 & 10 MHz and 1 PPS output, amplitude $\geq$ 0.5 V, stability after taming: $\leq$ $1\times10^{-11}$ \\
		\bottomrule
	\end{tabular}
\end{table*}

\textit{3) Electronic Switch.}
High-speed electronic switches are employed to enable time-division multiplexed MIMO measurements with  antenna arrays.
The custom-designed electronic switch is illustrated in Figs. \ref{device1} and \ref{block}.
Taking the 32-channel mmWave switch as an example, a cascaded control structure is adopted, where one single-pole four-throw (1SP4) module controls four single-pole eight-throw (1SP8) submodules. This compact architecture reduces hardware complexity while improving scalability. 
With interchangeable frequency-band submodules, the deployed 24--40 GHz switch remains operable over 6--40 GHz, with only a slight insertion-loss increase (from 10 dB to 13 dB at 15 GHz), which can be compensated by the system gain. 
Therefore, FR3 measurements can be supported without hardware modification.

\textit{4) Timing and synchronization modules.} 
Two GNSS-disciplined rubidium clocks are deployed at  Tx and Rx as the primary timing and frequency references for geographically separated dynamic measurements. Locked to the same Universal Coordinated Time (UTC) reference, they provide stable 10 MHz  and 1 PPS signals to maintain long-term Tx–Rx synchronization, with residual offsets compensated through system calibration. The integrated GNSS receiver further supports measurement logging and trajectory referencing.  
PXIe-6674T acts as a timing distribution module rather than an independent oscillator, forwarding   rubidium-clock reference and trigger signals to   PXIe modules through   PXI timing backplane for coherent acquisition and consistent timestamping.

\textit{5) Antenna arrays.} 
As shown in Fig. \ref{device1}, multiple antenna configurations are supported to accommodate different frequency bands and measurement scenarios. Together with the electronic switching network, the standardized RF and mechanical interfaces enable flexible spatial sampling, rapid antenna replacement, and reconfigurable multi-band channel measurements over the 10 MHz--44 GHz range.

\section{Visual Sensing Subsystem Design and Implementation}
This section presents the hardware design, processing workflow, and metrological characterization of the visual sensing subsystem, which measures the second core measurand of the system: spatio-temporally synchronized multi-modal environmental data. 
Through calibration and spatial registration with channel sounding subsystem, it provides accurate and synchronized environmental observations, distinguishing the proposed system from conventional standalone sensing devices.

\subsection{Overall Architecture}
The visual sensing subsystem provides a high-fidelity, spatio-temporally aligned digital representation of the measurement environment. 
In this work, the representation is used as a high-confidence environmental reference, rather than an unconditional ground truth for dynamic scenes, to support the interpretation of radio propagation phenomena, including dominant-scatterer identification, LoS/non-line-of-sight (NLoS) classification, and environment-specific feature extraction for channel modeling \cite{gong2025}.

The reliability of this environmental reference under dynamic conditions is ensured through joint spatial, temporal, and geometric constraints. First, the LiDAR, panoramic camera, GNSS antenna, and receiver antenna array are rigidly mounted on the same platform, providing a fixed spatial relationship among sensing and channel-measurement devices. Second, heterogeneous data streams are aligned using timestamps and frame indices, which enables channel snapshots, LiDAR frames, and camera images to be associated within a bounded temporal margin. Third, LiDAR-inertial processing and destaggering correction are applied to reduce motion- and scanning-induced geometric distortions in the point cloud. In this framework, the LiDAR point cloud  provides metric geometric constraints, while the panoramic camera supplies complementary visual texture and semantic context. Therefore, the visual sensing subsystem can provide reliable environmental sampling for static or quasi-static structures in dynamic measurement scenarios, while moving objects and ambiguous transient observations are not treated as absolute references.

\begin{figure}[!t]
	\centering
	\includegraphics[width=.45\textwidth]{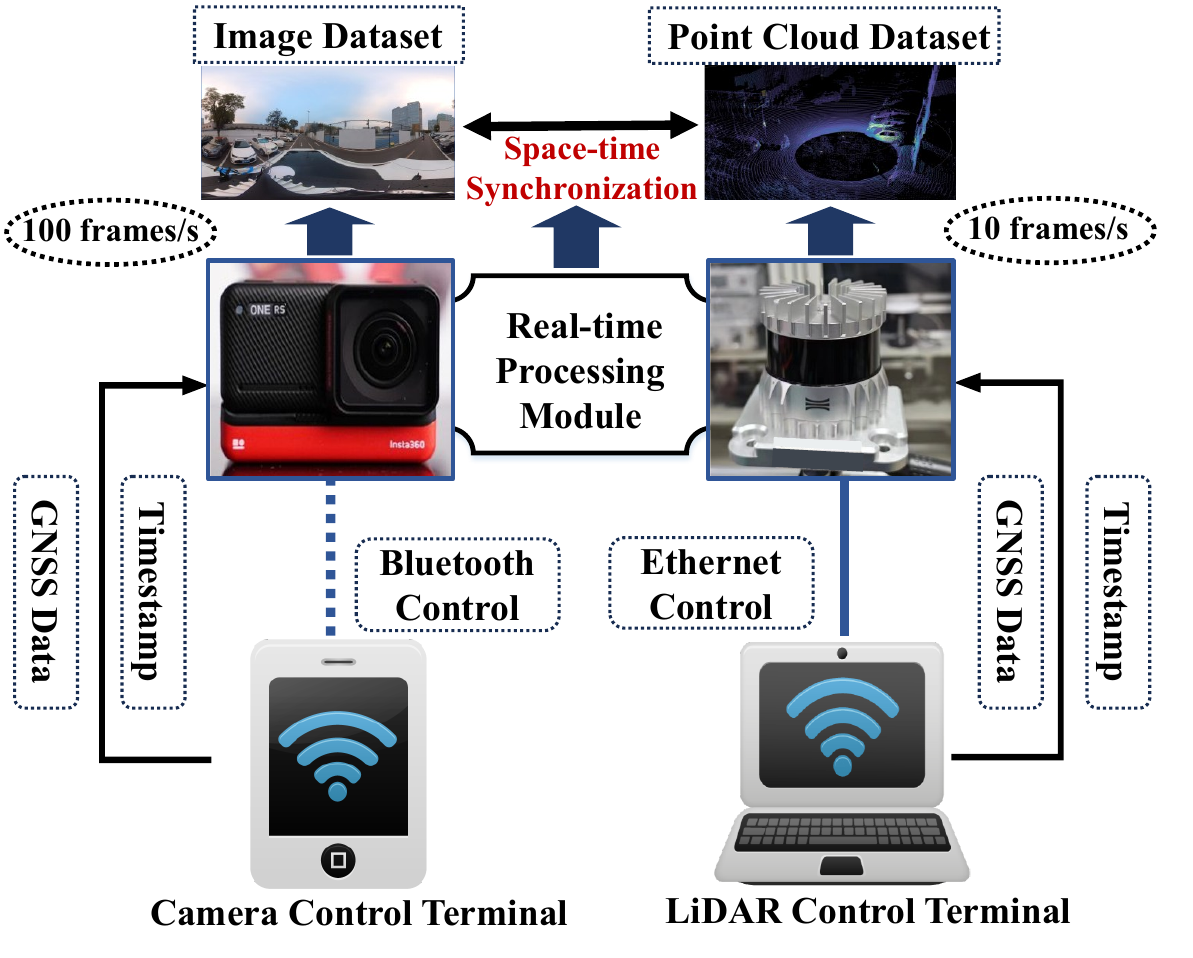} 
	\caption{System architecture of the environmental perception and reconstruction subsystem.}
	\label{fig:visual_arch}
\end{figure}

The general architecture of  visual sensing subsystem is depicted in Fig. \ref{fig:visual_arch}.
It is composed of hardware and software layers. 
The hardware layer consists of two main modules: a LiDAR scanning unit for 3D geometric reconstruction and a panoramic camera for visual data acquisition. The LiDAR is directly connected to and controlled by the host computer via Ethernet, whereas the camera is remotely operated through a Bluetooth-connected smartphone. As discussed in Section II, the smartphone serves only as an auxiliary control terminal for starting and stopping image acquisition and, when available, for embedding GNSS position and velocity metadata into the recorded visual data. These metadata are used solely for data annotation and post-processing convenience, rather than for system-level timing synchronization.

It should be emphasized that Bluetooth is used only for camera control and metadata exchange. The image data are stored locally on the camera and transferred offline after the measurement campaign. Therefore, the Bluetooth link does not carry high-rate sensing data and does not constitute a bottleneck under high-mobility or high-data-rate measurement scenarios. The multi-modal synchronization among channel, LiDAR, and camera data follows the timestamp-based post-processing alignment strategy described in Section II.

The software layer, primarily built upon the Robot Operating System (ROS), manages hardware drivers, data acquisition, and real-time processing. 
Raw data streams, including point clouds from LiDAR, RGB images from the camera, and pose information (position and orientation), are collected and pre-processed. 
These processed data form a digital representation of the environment.
This dataset is then used for tasks such as instance segmentation, scatterer identification, and estimation of physical and electromagnetic parameters of  objects.

\subsection{Hardware Configuration}
To achieve high-precision and robust environment reconstruction, we have selected state-of-the-art sensing hardware and implemented a tightly-coupled LiDAR-inertial odometry framework for simultaneous localization and mapping (SLAM).
The core components of the visual sensing subsystem are a high-resolution LiDAR and a versatile panoramic camera. 
Their key specifications are summarized in Table \ref{performance}.

The LiDAR scanning module is built around the OUSTER OS1-128-REV7, a 128-beam LiDAR sensor. 
It offers a 360° horizontal and 45° vertical field of view, enabling comprehensive environmental coverage. 
It should be noted that the  IMU is integrated into the LiDAR and therefore shares the same physical housing and timestamping framework as the LiDAR measurements. 
The IMU data are utilized for motion compensation and pose estimation during dynamic measurements.
With a measurement accuracy of ±5 cm and a range of up to 200 meters, it can rapidly and precisely capture the geometry of the surrounding environment. 
The module is integrated with a high-performance processing unit and a portable display, allowing for real-time visualization, multi-frame registration, and loop closure detection of the point cloud data \cite{QI2023241}. 
This immediate feedback ensures the validity and utility of the collected data during measurement campaigns.
The video and image  acquisition module utilizes the Insta360 ONE RS, a modular camera system. 
Its 4K Wide Angle Lens, featuring a 1/2-inch 48MP complementary metal-oxide-semiconductor (CMOS) sensor, supports up to 4K photos. 
For dynamic measurement scenarios, an active high dynamic range (HDR) mode combined with FlowState stabilization ensures high-quality, stable footage.


\subsection{3D Environment Reconstruction Implementation}
To construct a globally consistent map of the environment, we employ a tightly-coupled LiDAR-inertial SLAM algorithm. 
This approach fuses high-frequency inertial measurements from the IMU with lower-frequency, feature-rich LiDAR scans to achieve robust and accurate state estimation \cite{Nishio2021}.



The state vector $\mathbf{x}$ to be estimated includes the  rotation $\mathbf{R}$, position $\mathbf{p}$, velocity $\mathbf{v}$, and the IMU biases for the gyroscope $\mathbf{b}_g$ and accelerometer $\mathbf{b}_a$. The state propagation is driven by the IMU measurements. Between two consecutive LiDAR scans, the system state is predicted forward in time using the IMU kinematic model:
\begin{align}
	\dot{\mathbf{R}}(t) &= \mathbf{R}(t) [\boldsymbol{\omega}_m(t) - \mathbf{b}_g(t)]_{\times} \\
	\dot{\mathbf{p}}(t) &= \mathbf{v}(t) \\
	\dot{\mathbf{v}}(t) &= \mathbf{R}(t)(\mathbf{a}_m(t) - \mathbf{b}_a(t)) + \mathbf{g},
\end{align}
where $[\cdot]_{\times}$ is the skew-symmetric matrix operator, $\boldsymbol{\omega}_m(t)$ and $\mathbf{a}_m(t)$ are the measured angular velocity and acceleration,  $\mathbf{g}$ is the gravity vector, and the dot operator represents the time derivative.

Upon the arrival of a new LiDAR scan, the points in the scan are used to form measurement residuals. These residuals quantify the discrepancy between  measured points and  existing map, which is built from previous scans. 
The core of the algorithm is to minimize a cost function that incorporates both   IMU propagation errors and   LiDAR measurement residuals:
\begin{equation}
	\min_{\mathbf{x}} \left\{ \sum_{i \in \mathcal{I}} \| \mathbf{r}_{\mathcal{I}}(\hat{\mathbf{z}}_i, \mathbf{x}) \|^2_{\mathbf{P}_i^{-1}} + \sum_{j \in \mathcal{L}} \| \mathbf{r}_{\mathcal{L}}(\hat{\mathbf{z}}_j, \mathbf{x}) \|^2_{\mathbf{Q}_j^{-1}} \right\},
\end{equation}
where $\mathbf{r}_{\mathcal{I}}$ and $\mathbf{r}_{\mathcal{L}}$ represent the residual functions for  IMU and LiDAR measurements, respectively, with $\mathbf{P}_{i}$ and $\mathbf{Q}_{j}$ being their corresponding covariance matrices;
\(\mathcal{I}\) and \(\mathcal{L}\) denote the index sets of IMU and LiDAR measurements, respectively, \(\hat{\mathbf{z}}_i\) and \(\hat{\mathbf{z}}_j\) are the corresponding preprocessed measurements.
The notation \(\|\mathbf{\cdot}\|^2_{\mathbf{M}}\) denotes the squared Mahalanobis norm.
The nonlinear least-squares problem is solved iteratively, with LiDAR-derived corrections used to refine the state trajectory and update the IMU biases. To ensure real-time performance, the global map is maintained using an incremental k-dimensional (k-d) tree, which supports efficient map updates, nearest-neighbor search, residual computation, and loop-closure detection. Processing the full measurement dataset then yields a high-resolution, globally consistent 3D point cloud map as a digital representation of the measurement environment.

\subsection{Visual Data Fusion for Real-Time Sensing}

While the global 3D map provides a static environmental representation, real-time sensing is necessary to capture dynamic scene variations and associate instantaneous channel responses with physical objects. 
To this end, synchronized LiDAR point clouds and camera images are fused, integrating image-level semantics with metric 3D geometry.

\begin{figure}[!t]
	\centering
	\includegraphics[width=0.45\textwidth]{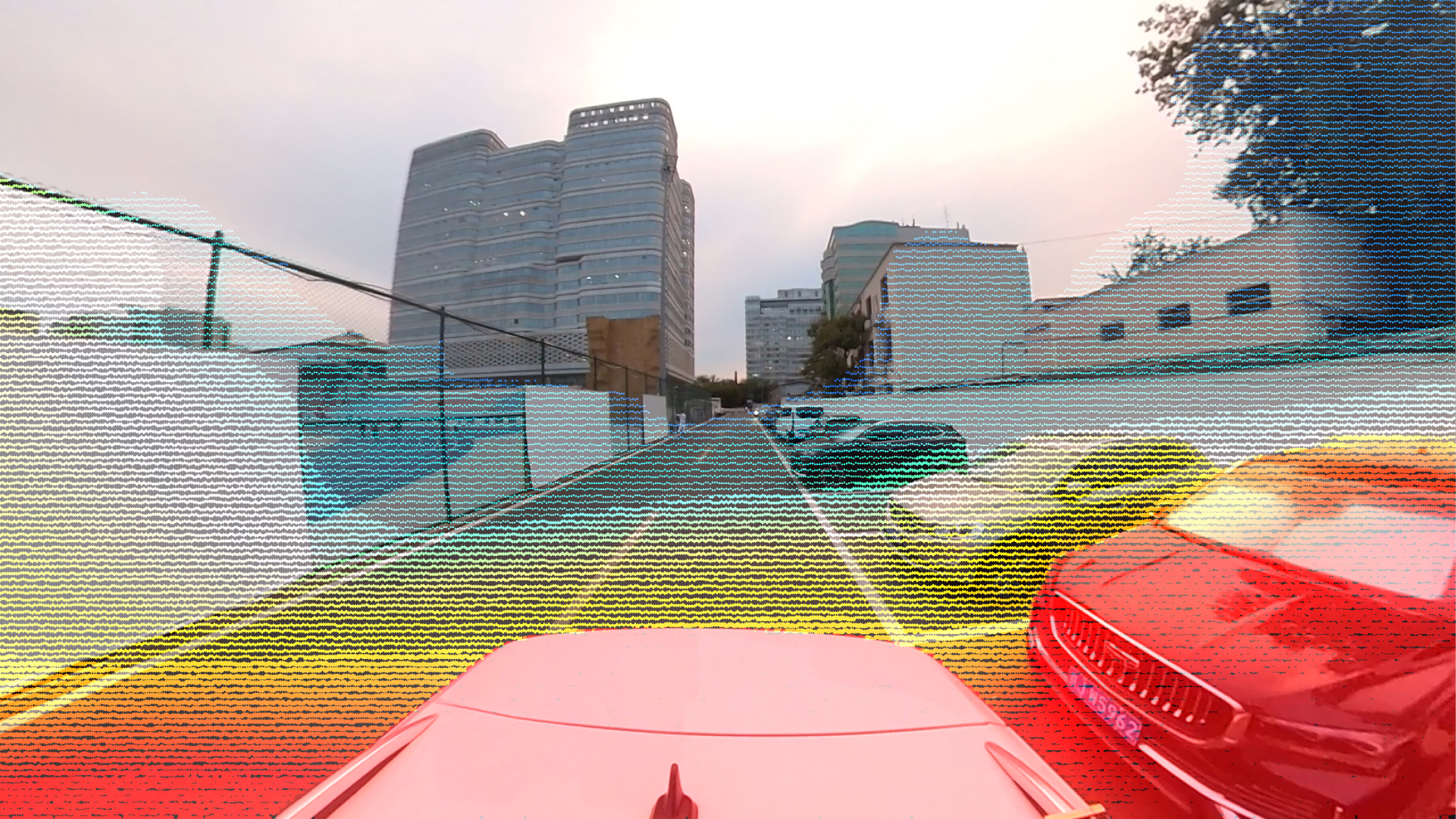} 
	\caption{Example of fusing a real-time LiDAR point cloud frame with a synchronized camera image. The points are colored by distance, providing depth perception overlaid on the visual scene.}
	\label{fig:fusion_example}
\end{figure}

The core of the fusion process is the projection of 3D points from the LiDAR coordinate system onto the 2D image plane. This requires precise extrinsic and intrinsic calibration. 
Let $\mathbf{P}_L = [X_L, Y_L, Z_L]^T$ be 3D point coordinates in the LiDAR's coordinate frame. First, it is transformed into the camera's coordinate frame $\mathbf{P}_C = [X_C, Y_C, Z_C]^T$ using the rigid body transformation matrix $\mathbf{T}_{C \leftarrow L}$, which represents the rotation $\mathbf{R}_{C \leftarrow L}$ and translation $\mathbf{t}_{C \leftarrow L}$ from LiDAR to camera:
\begin{equation}
	\begin{bmatrix} \mathbf{P}_C \\ 1 \end{bmatrix} = \mathbf{T}_{C \leftarrow L} \begin{bmatrix} \mathbf{P}_L \\ 1 \end{bmatrix} = \begin{bmatrix} \mathbf{R}_{C \leftarrow L} & \mathbf{t}_{C \leftarrow L} \\ \mathbf{0}^T & 1 \end{bmatrix} \begin{bmatrix} \mathbf{P}_L \\ 1 \end{bmatrix}.
\end{equation}
Next, the 3D point $\mathbf{P}_C$ is projected onto the image plane to obtain the pixel coordinates $\mathbf{p} = [u, v]^T$ using the camera's intrinsic matrix $\mathbf{K}$:
\begin{equation}
	Z_C \begin{bmatrix} u \\ v \\ 1 \end{bmatrix} = \mathbf{K} \mathbf{P}_C = \begin{bmatrix} f_x & 0 & c_x \\ 0 & f_y & c_y \\ 0 & 0 & 1 \end{bmatrix} \begin{bmatrix} X_C \\ Y_C \\ Z_C \end{bmatrix},
\end{equation}
where $(f_x, f_y)$ are the focal lengths and $(c_x, c_y)$ is the principal point of the camera.
As illustrated in Fig.~\ref{fig:fusion_example}, the calibrated projection accurately overlays LiDAR points onto the corresponding RGB image. With the point cloud color-coded by range, the 3D scene structure, including the ground plane, buildings, fences, and vehicles, can be directly interpreted in the image domain. This fusion links image-derived semantic labels to metric 3D point clusters, enabling the identification of major reflectors and scatterers by both type and location. Such multi-modal context provides the basis for subsequent channel--environment joint analysis.

\section{System Verification}
This section provides a quantitative verification of the proposed measurement system from both channel sounding and environmental sensing perspectives. The verification focuses on the key metrological characteristics required for environment-aware channel measurement, and quantitatively evaluates the measurement accuracy, stability, and synchronization performance of the system, thereby ensuring that the generated data satisfy the metrological requirements for 6G wireless communication research.

\subsection{System Calibration}

\begin{figure}[!t]
	\centering
	\includegraphics[width=.48\textwidth]{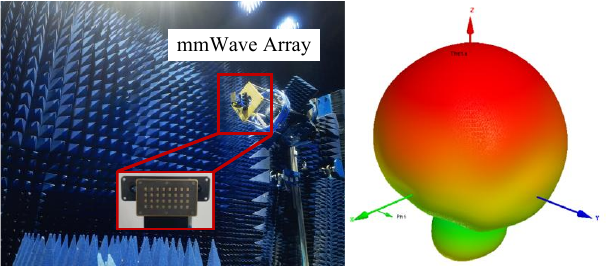}%
	\caption{Anechoic chamber measurement setup for characterization of the per-element gain and phase patterns of mmWave antenna array.
	}                    
	\label{antenna}
\end{figure}

\begin{figure}[!t]
	\centering
	\includegraphics[width=.48\textwidth]{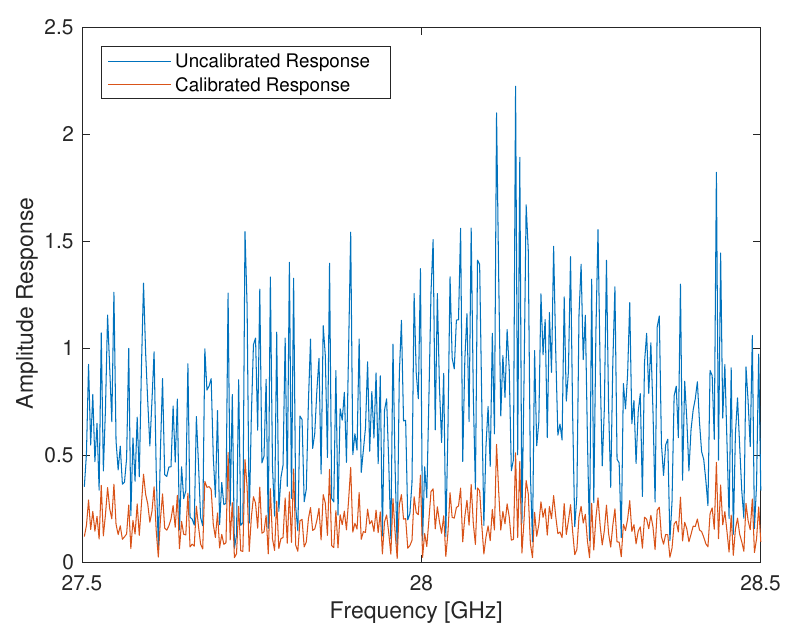}%
	\caption{Calibrated and uncalibrated amplitude responses of the system.
	}                    
	\label{Cal}
\end{figure}

System calibration is crucial to ensure accuracy, repeatability, and spatial fidelity in channel measurements. 
The procedure consists of three stages: instrument self-calibration, back-to-back (B2B) calibration, and antenna calibration \cite{bae2025}. 
Instrument self-calibration compensates for hardware drift, temperature variation, and aging, with automated routines adjusting frequency references  and modulators/demodulators before each measurement. 
B2B calibration eliminates system distortions by directly connecting Tx and Rx through attenuated RF cables, extracting the intrinsic system response for data compensation. 
Antenna calibration corrects directionality, gain, and polarization effects. 
Radiation patterns are measured in an anechoic chamber using azimuthal rotation, as shown in Fig. \ref{antenna}, and the results are used to normalize antenna gain and construct accurate steering vectors, thereby improving angle-domain parameter estimation. 
It should be noted that the anechoic-chamber calibration characterizes the intrinsic gain, phase response, and array manifold of the antenna system, which are independent of the propagation environment and therefore remain valid in practical multipath scenarios.
The residual calibration uncertainty is below $2^\circ$ in phase and 1 dB in gain across the operating bandwidth. 
Error propagation analysis indicates that the resulting uncertainties are below 0.1 ns for delay spread, below $1^\circ$ for angular spread, and below 1 dB for path loss estimation. 
These values are significantly smaller than the corresponding system resolutions and measurement ranges, confirming the adequacy of the calibration procedure for accurate channel characterization in real-world environments.
Fig. \ref{Cal} shows the calibrated and uncalibrated amplitude responses, demonstrating that  calibration process effectively eliminates   frequency-dependent gain variations and smooths the overall response.

\subsection{Dynamic Range}

\begin{figure}[!t]
	\centering
	
	\includegraphics[width=.45\textwidth]{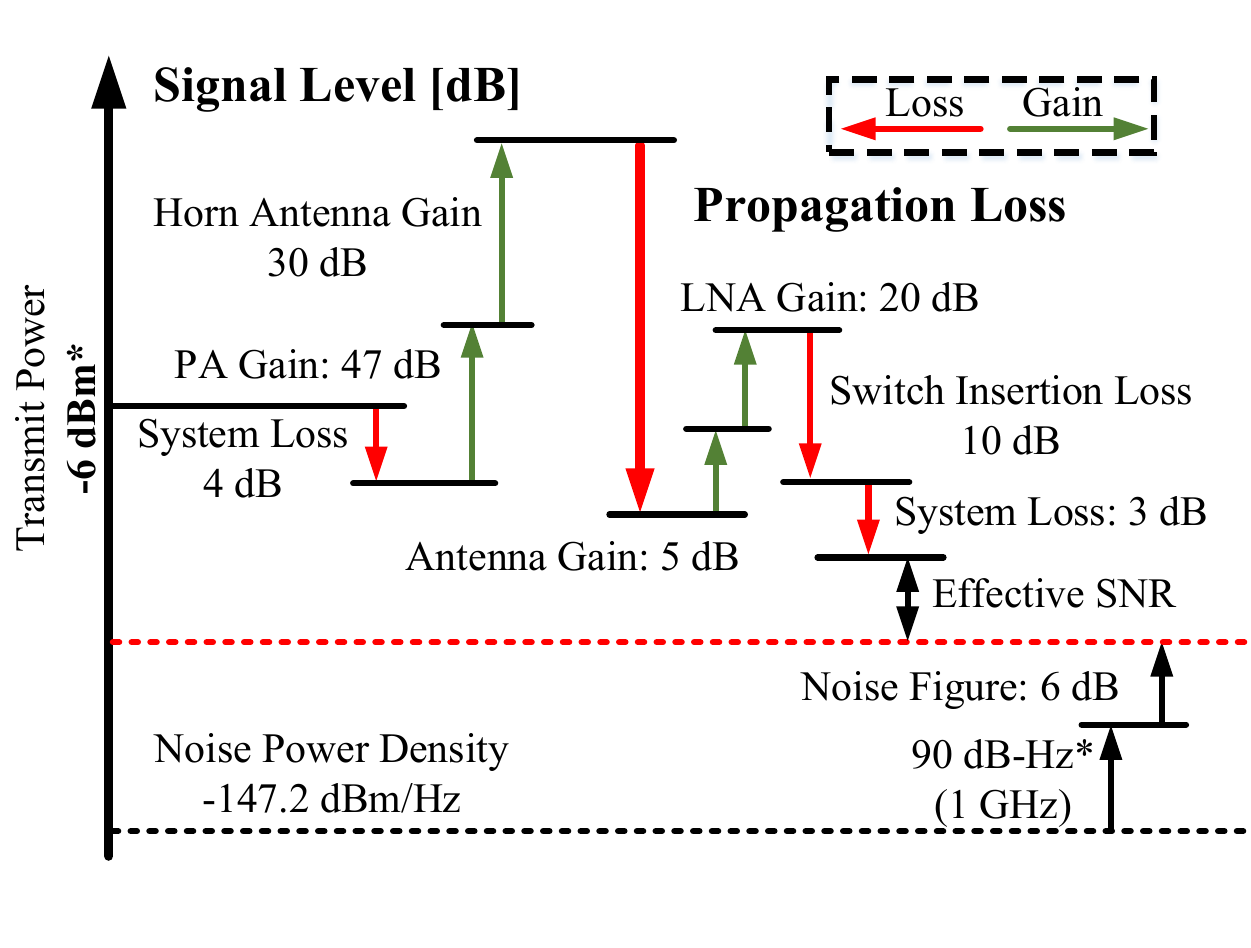}%
	
	\caption{
   Example of a link budget analysis for the channel sounder anchored at 28 GHz using a high-gain narrow-beam horn as the transmit antenna and a 32-element rectangular array at the receiver.
    ``*" indicates adjustable value. 
	}
	\label{link}
\end{figure}

Fig. \ref{link} presents the link-budget-based dynamic range evaluation of the 28 GHz channel sounder. 
The maximum measurable propagation loss is defined as the largest path loss under which the effective received SNR remains above the detection threshold. 
In the evaluation, Tx is configured at its maximum link-budget setting, consisting of a narrow-beam horn antenna with a gain of 30 dB and  a PA with a gain of  47 dB.
Rx employs a 32-element rectangular array with an average gain of 5 dB, followed by a LNA with a gain of 20 dB and a high-speed electronic switch with an  insertion loss of 10 dB.
The additional cable and interface losses at the Tx and Rx sides are approximately   4 dB and 3 dB, respectively.
The received signal power after the RF front-end can be expressed as
\begin{equation}
	P_{\mathrm{rx}} =
	P_{\mathrm{tx}} + G_{\mathrm{tx}} + G_{\mathrm{rx}} + G_{\mathrm{LNA}} 
	- L_{\mathrm{tx}} - L_{\mathrm{rx}} - L_{\mathrm{sw}} - L,
\end{equation}
where $P_{\mathrm{tx}}$ is the output power of Tx and power amplifier chain, $G_{\mathrm{tx}}$ and $G_{\mathrm{rx}}$ denote the Tx and Rx antenna gains, $L_{\mathrm{tx}}$ and $L_{\mathrm{rx}}$ are the Tx/Rx cable and interface losses, $L_{\mathrm{sw}}$ is the switch insertion loss, $G_{\mathrm{LNA}}$ is the LNA gain, and $L$ is the propagation loss to be measured.

The corresponding noise power is calculated as
\begin{equation}
	P_{\mathrm{n}} = N_0 + 10\log_{10}(B) + F_{\mathrm{rx}},
\end{equation}
where $N_0$ is -147.2 dBm/Hz, which is the measured noise spectral density of the PXIe-5775 receiver, $B$ is the effective measurement bandwidth, and $F_{\mathrm{rx}}\approx6$  dB is the cascaded receiver noise figure. 

By imposing the minimum detectable SNR threshold $\gamma_{\min}=0$ dB, the maximum measurable propagation loss is obtained as
\begin{equation}
	L_{\max}=P_{\mathrm{rx}}-P_{\mathrm{n}}-\gamma_{\min},
\end{equation}
Substituting the parameters in Fig. \ref{link}, the dynamic range of the 28 GHz sounder is approximately 130 dB. 
This value is further supported by the measurement results in Section VI, where the maximum observed path loss is approximately 122 dB, leaving a sufficient measurement margin of about 8 dB. 
Therefore, the proposed platform provides adequate dynamic range for reliable wideband channel measurements in the considered dynamic scenarios.

\subsection{Phase Stability}
The stability of phase information plays a crucial role in multipath channel measurement and modeling. Any significant phase drift in the measurement system degrades delay resolution, distorts multipath parameter estimation, and causes inconsistencies among measurements over time, thereby reducing the reliability of the derived channel model. To ensure measurement accuracy, the long-term phase stability of the channel sounder is evaluated.

 \begin{figure}[!t]
	\centering
	\includegraphics[width=.48\textwidth]{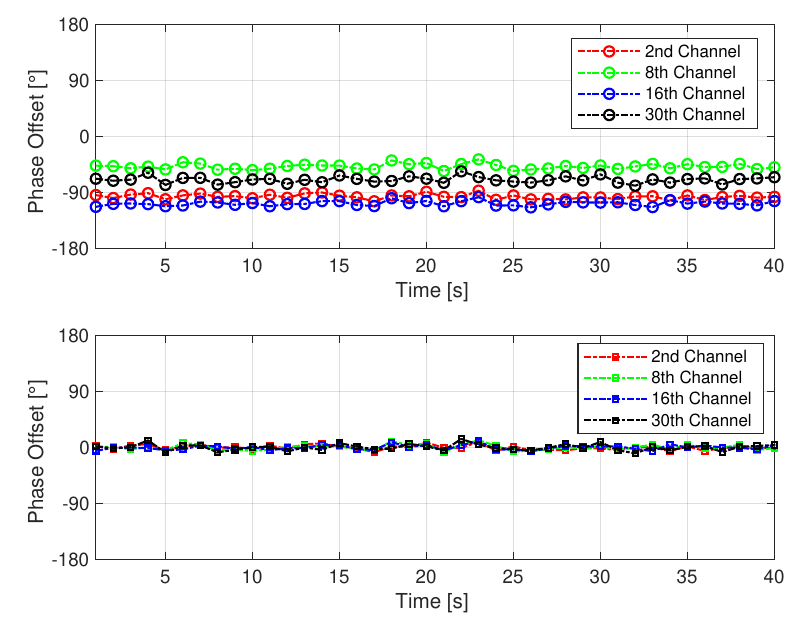}%
	\caption{Calibrated and uncalibrated phase offsets with respect to Channel 1.
	}                    
	\label{phase}
\end{figure}

The experiment is conducted in a static indoor environment using 28 GHz as an example. 
In SIMO measurements, electronic switch introduces fixed phase offsets across different channels, which should remain stable over time to ensure system consistency.
Fig. \ref{phase} presents the phase stability assessment of the proposed channel sounding system during a representative measurement interval.
The upper plot presents the uncalibrated phase offsets of multiple channels relative to Channel~1, where noticeable static offsets and random fluctuations can be observed. After applying calibration, as illustrated in the lower plot, these offsets are effectively compensated, and all channels exhibit highly consistent and stable phase responses with negligible drift over time.
This demonstrates the stability of the RF front-end, synchronization network, and data acquisition system. 

\begin{figure}[!t]
	\centering
	\includegraphics[width=.5\textwidth]{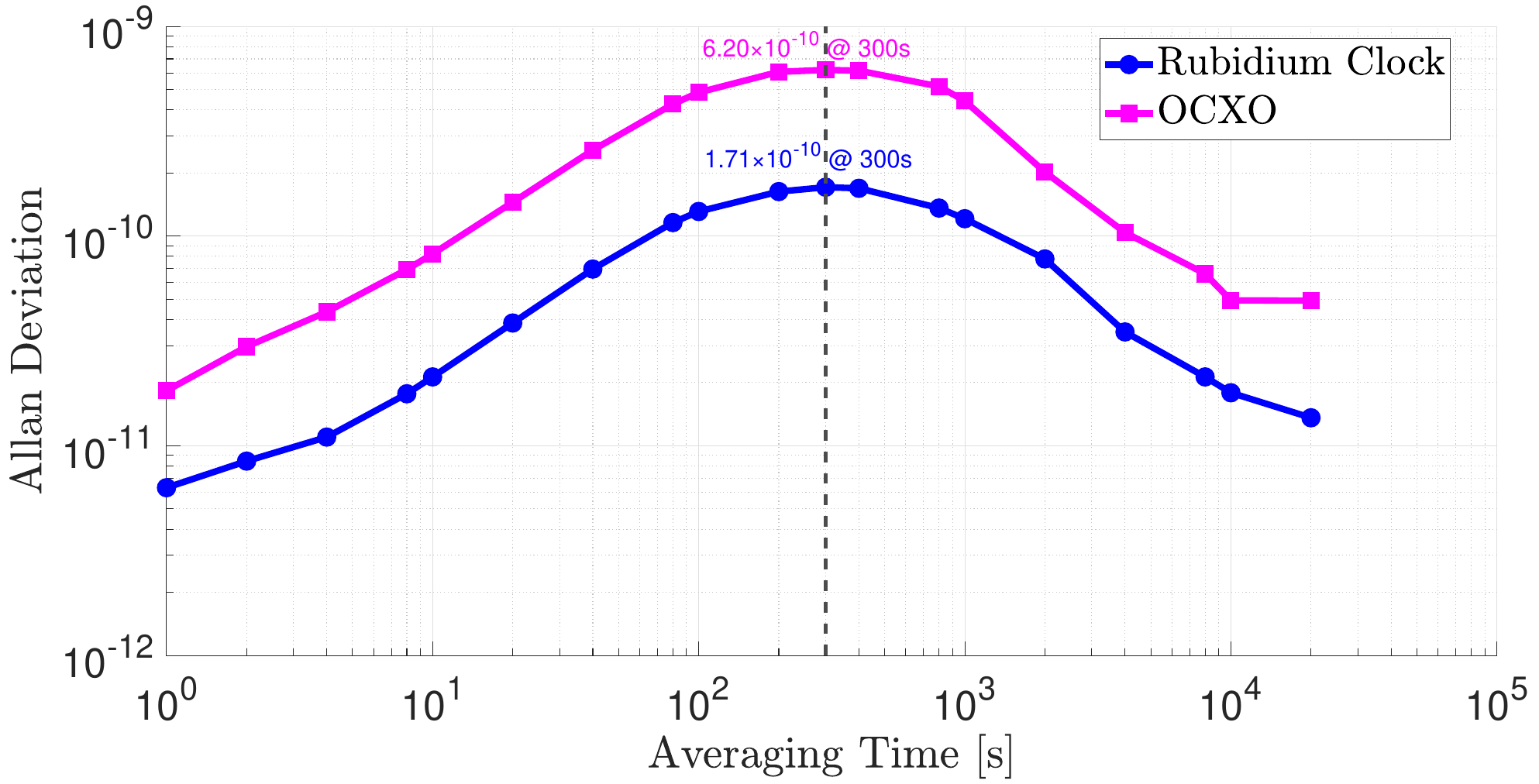}%
	\caption{Experimental comparison of   Allan deviation of the 10 MHz outputs of   rubidium oscillator and OCXO.
	}                    
	\label{clock}
\end{figure}

Furthermore, the proposed system employs a GNSS-disciplined rubidium clock as the common reference source. 
Fig. \ref{clock} presents an experimental Allan-deviation comparison between the 10 MHz outputs of  rubidium oscillator and oven-controlled crystal oscillator (OCXO), conducted by using a Microchip 53100A frequency-stability analyzer. 
The measurement durations are approximately 12 h 50 min for rubidium oscillator and 17 h 3 min for  OCXO. 
At an averaging time of 300 s, their Allan deviations are $1.71\times10^{-10}$ and $6.20\times10^{-10}$, respectively. 
Moreover, rubidium clock continues to improve as the averaging time increases, eventually approaching the $10^{-11}$ level. 
Notably, the non-monotonic behavior at intermediate averaging times originates from a residual initial frequency offset and the associated settling transient after switching the measurement interface. 
Therefore, these curves are interpreted as an empirical comparison under the specific test conditions, rather than as a canonical noise-type decomposition of free-running oscillators. 
Nevertheless, the consistently lower Allan deviation of the rubidium oscillator supports its use as the stable frequency reference for phase-coherent channel sounding.

\subsection{LiDAR Data Destaggering}

\begin{figure}[!t]
	\centering
	\subfloat[]{\includegraphics[width=.45\textwidth]{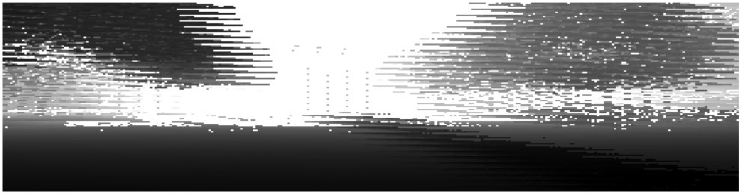}%
		\label{lidar-a}}
        
	\subfloat[]{\includegraphics[width=.45\textwidth]{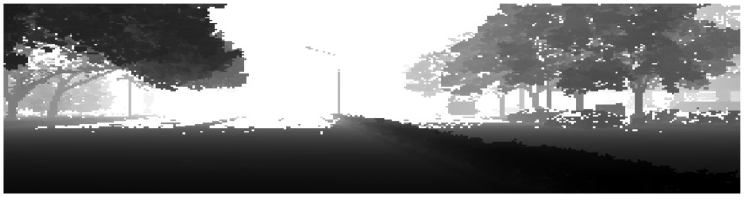}%
		\label{lidar-b}}
	\caption{
	Comparison of range grayscale images before and after destaggering correction. (a) Before destaggering. (b)  After destaggering. 
    Both images share a common grayscale map and display range, clamped to the 1st--99th percentile range. Zero or invalid ranges are masked in white.
	}
	\label{fig:destagger_pair}
\end{figure}

As an active sensing module, LiDAR provides high-resolution 3D geometric information for environment reconstruction. However, raw LiDAR frames may contain staggered geometric artifacts and therefore cannot be directly used for high-level perception tasks such as object detection, edge extraction, and 3D mapping.
As shown in Fig. \ref{fig:destagger_pair}(a), the uncorrected range image exhibits evident distortion, where vertical structures such as poles, trees, and building edges appear tilted or discontinuous. Such distortion may lead to inaccurate feature extraction and degraded localization and mapping performance; therefore, geometric correction is an essential preprocessing step.

To quantify the correction effect, static vertical structures in the scene are used as reference features. The angular deviation of these structures from the vertical direction is evaluated before and after destaggering.
Before correction, the average angular error of vertical structures (e.g., poles, buildings) is $1.2^{\circ} \pm 0.3^{\circ}$, corresponding to a cross-range position error of approximately 4.2 m at a distance of 200 m, calculated by 
\begin{equation}
	e_{\perp}=R\tan(e_{\theta}),
\end{equation}
where $e_{\perp}$ denotes the cross-range position error caused by the angular deviation, $R$ is the distance between LiDAR and the observed object, and $e_{\theta}$ represents the angular error of the observed structure with respect to its reference direction.
After destaggering, the angular error is reduced to $0.08^{\circ} \pm 0.02^{\circ}$, corresponding to a cross-range error of approximately 0.28 m at the same distance.
This represents a 15 times improvement in geometric consistency.

The staggering artifact mainly originates from the row-dependent timing offsets among different LiDAR firing channels. Since the laser channels corresponding to different image rows are not sampled at exactly the same azimuth angle, vertical structures may appear tilted or discontinuous in the raw range image. This geometric distortion degrades the reliability of subsequent perception tasks, such as edge extraction, object localization, and 3D environment reconstruction.

Destaggering aims to compensate for this row-wise azimuthal displacement and restore the geometric consistency of the range image. 
The horizontal offset $s_r$ provided in the LiDAR metadata represents the pre-calibrated shift of the $r$-th row caused by internal firing delay. 
Therefore, the correction can be implemented as an independent circular shift along each row. 
Let $S\in\mathbb{R}^{H\times W}$ and $D\in\mathbb{R}^{H\times W}$ denote the staggered and corrected range frames, respectively, where $H$ is the number of vertical channels and $W$ is the number of columns. 
The destaggering operation is given by
\begin{equation}
D[r,c] \;=\; S\!\big[r,\,(c+s_r)\bmod W\big].
\label{eq:destagger-2d}
\end{equation}
Here, the pixel at $[r,c]$ in the corrected image $D$ is taken from the same row in $S$, shifted by $s_r$ columns. 
The modulo operation ensures circular wrapping, which is essential for \(360^{\circ}\) panoramic scans.
As shown in Fig. \ref{fig:destagger_pair}(b), the corrected range image exhibits substantially improved alignment of vertical structures, indicating that the proposed destaggering procedure effectively suppresses scanning-induced geometric distortion. This correction improves the geometric reliability of the LiDAR data and provides a more accurate basis for subsequent environment reconstruction and multi-modal feature association.

\begin{figure}[!t]
	\centering
	
	\subfloat[]{\includegraphics[width=.45\textwidth]{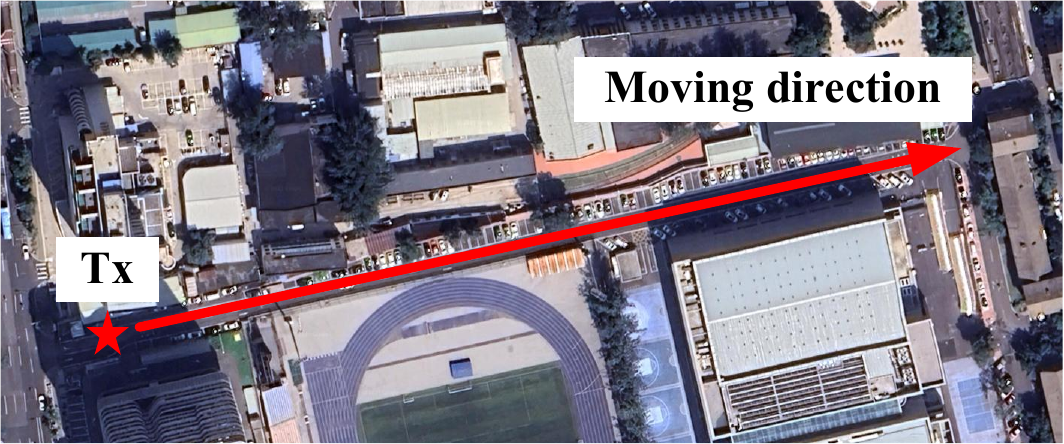}%
		\label{environment}}
	
	\subfloat[]{\includegraphics[width=.45\textwidth]{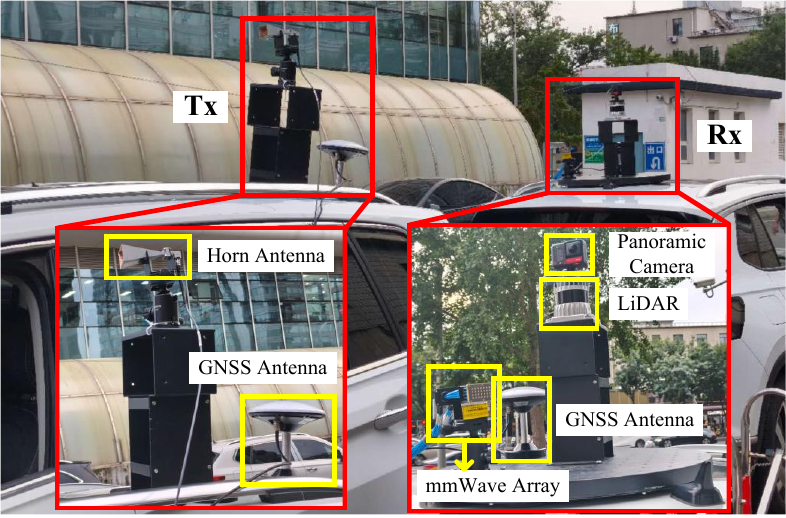}%
		\label{device3}}
	
	\caption{
	(a) Measurement environment;
	(b) Measurement equipment.
	}
	\label{measurement}
\end{figure}

\section{Sample Measurement}
This section presents a representative 28 GHz V2I campaign to validate the proposed system under dynamic operating conditions. Instead of deriving a site-specific channel model, the campaign focuses on verifying the core measurement capabilities, including large-scale attenuation characterization, delay- and angular-domain multipath resolution, and multi-modal association of dominant channel clusters with physical environmental features.

\subsection{Measurement Campaign}
To validate the   proposed platform,  a 28 GHz SIMO V2I  measurement campaign with   1 GHz bandwidth is conducted in a representative urban scenario, as illustrated in Fig. \ref{measurement}(a). 
The   route is an approximately 400 m straight road, with low walls and parked vehicles on the left side and a low fence on the right.
Apart from a few pedestrians and passing vehicles, no major obstruction is present.
Fig. \ref{measurement}(b) shows  the test vehicles and measurement setup. 
Tx employs a narrow-beam horn antenna with an effective beamwidth of $12^{\circ}–19^{\circ}$, while Rx is equipped with a 32-element array antenna. 
A custom magnetic mount rigidly integrates the array, LiDAR, panoramic camera, and GNSS antenna on the same platform, enabling synchronized multi-modal data acquisition during mobility. Detailed hardware specifications are listed in Table  \ref{hardware}.
To maintain LoS conditions, two identical vehicles are deployed, with Tx vehicle kept stationary and Rx vehicle moving along the predefined route at an average speed of 20 km/h. 
Rx collects SIMO channel data at   50 snapshots/s, and the  measurement procedure is repeated five times to ensure sufficient data for performance validation.

\subsection{Path Loss Verification}

 \begin{figure}[!t]
	\centering
	\includegraphics[width=.45\textwidth]{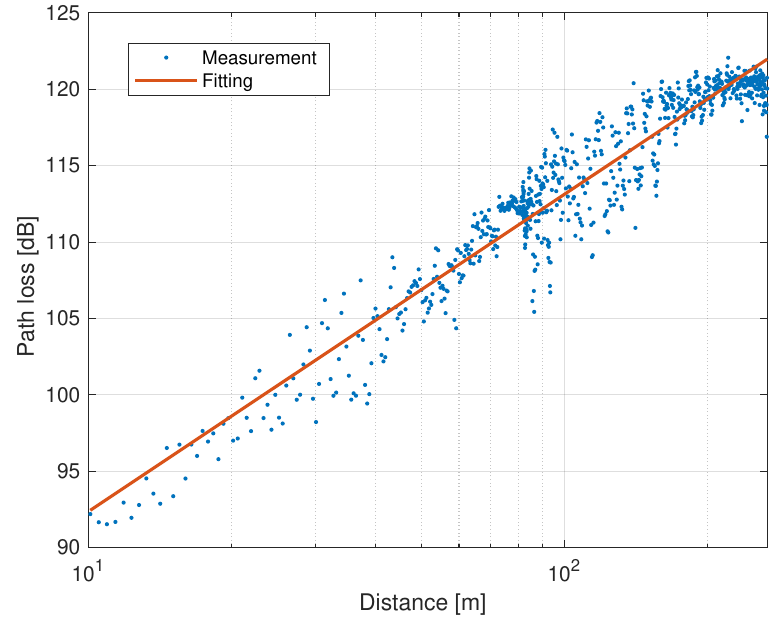}%
	\caption{The measured PL against Tx–Rx separation distance.
	}                    
	\label{PL}
\end{figure}

Path loss (PL) is a representative large-scale propagation characteristic that quantifies the attenuation of radio waves over long distances, which can be estimated from the average channel gain as follows:
\begin{equation}
	\label{math2}
	PL =  - 10{\log _{10}}(\frac{1}{W}\sum\limits_{T = t - \frac{W}{2}}^{t + \frac{W}{2} - 1} {\sum\limits_{\tau  = 1}^{{N_f} } {{{\left| {h(T,\tau )} \right|}^2}} } ),
\end{equation}
where \(PL\) denotes path loss in dB, \(h(T,\tau)\) is the calibrated channel impulse response at snapshot \(T\) and delay bin \(\tau\), \(t\) is the center snapshot index of the sliding window, \(T\) is the snapshot index within the window, \(N_f\) is the number of delay bins, and \(W\) denotes the window length corresponding to a travel distance of \(40\lambda\), with \(\lambda\) being the wavelength.
Fig. \ref{PL} presents the measured PL against Tx–Rx separation distance, where the blue points denote individual measurements, the red line is log-distance fitting. It can be observed that the measured PL closely follows the log-distance model, with a path loss exponent of approximately 2.07. 
It indicates that the system can accurately capture the large-scale attenuation trend. The dynamic range of the system allows reliable measurements up to 280 m, with the maximum observed path loss around 122 dB. Compared with the theoretical measurable range of 130 dB derived from link budget, this confirms that the platform maintains sufficient SNR margin across the entire measurement route.

\subsection{Power Delay Profile and Multipath Evolution}

\begin{figure}[!t]
	\centering
	
	\subfloat[]{\includegraphics[width=.45\textwidth]{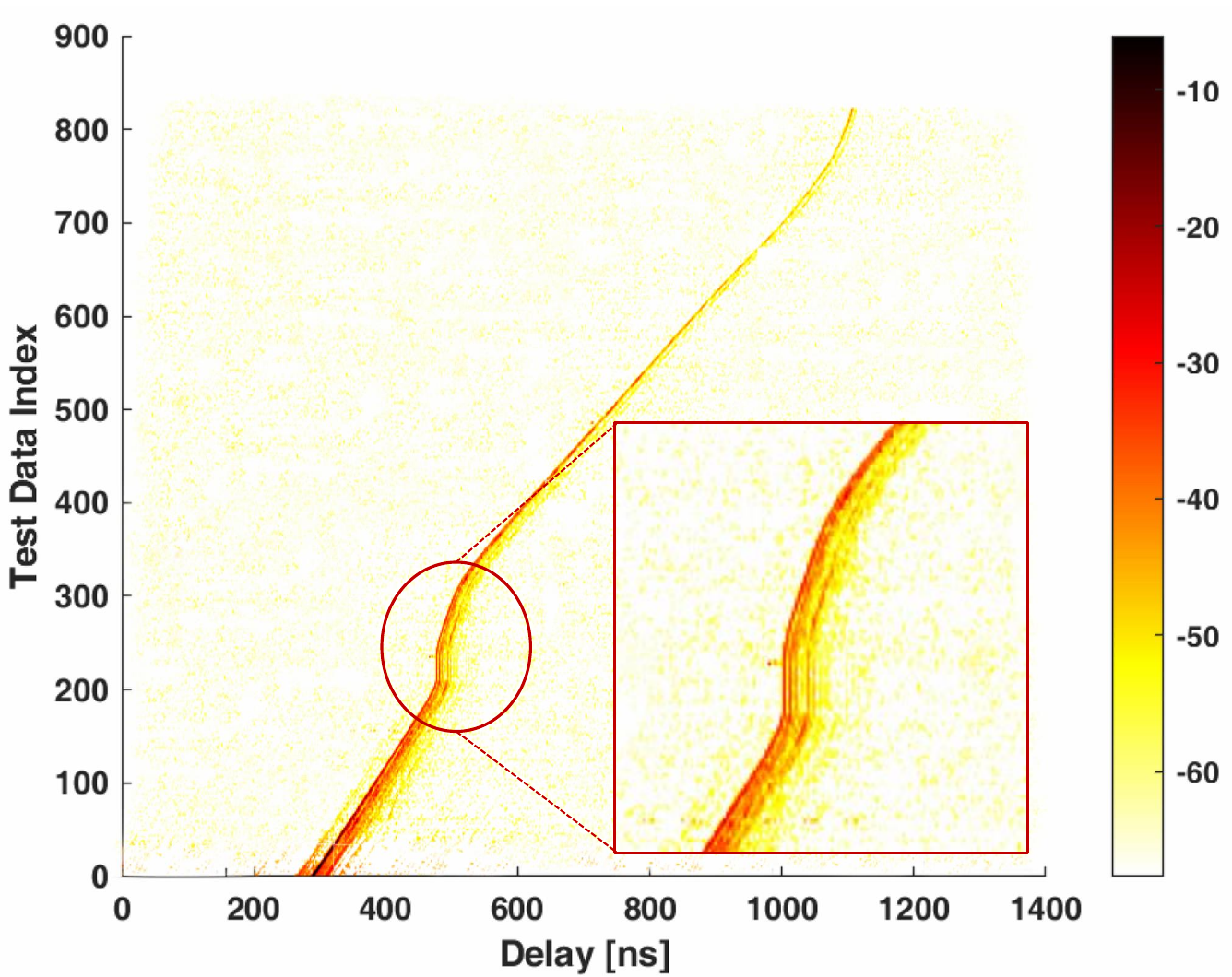}%
		\label{pdp}}
		\quad
	\subfloat[]{\includegraphics[width=.45\textwidth]{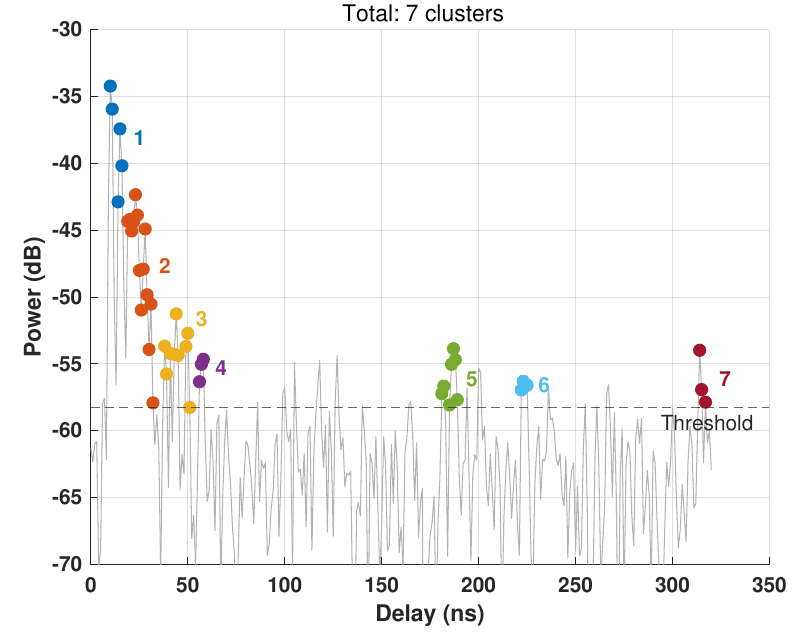}%
		\label{cluster}}
		\quad
	\subfloat[]{\includegraphics[width=.45\textwidth]{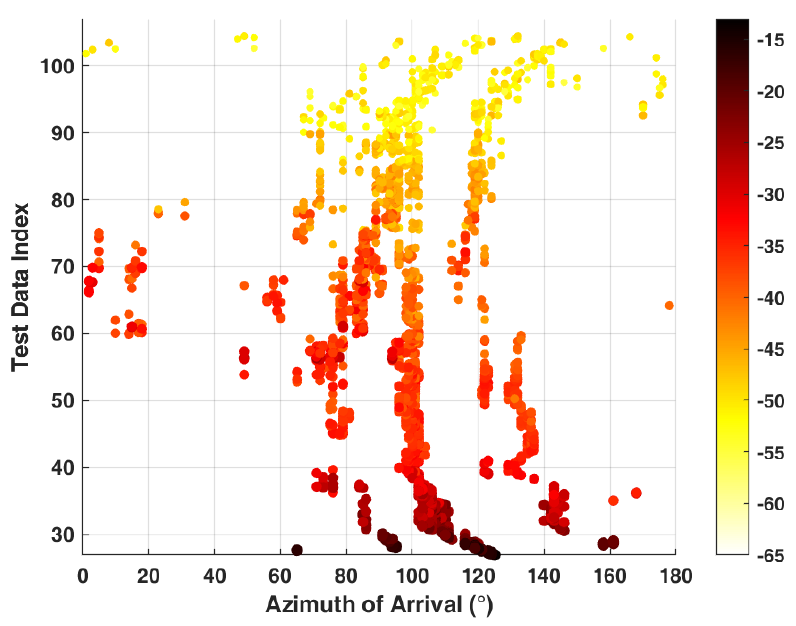}%
		\label{sage}}
	
	\caption{
	(a) The PDP evolution over the measurement campaign;
	(b) Multipath clustering results based on the measured PDP;
    (c) SAGE-based angle-of-arrival estimation results over the measurement campaign.
	}
	\label{multipath}
\end{figure}

\begin{figure*}[!t]
	\centering
	\includegraphics[width=1\textwidth]{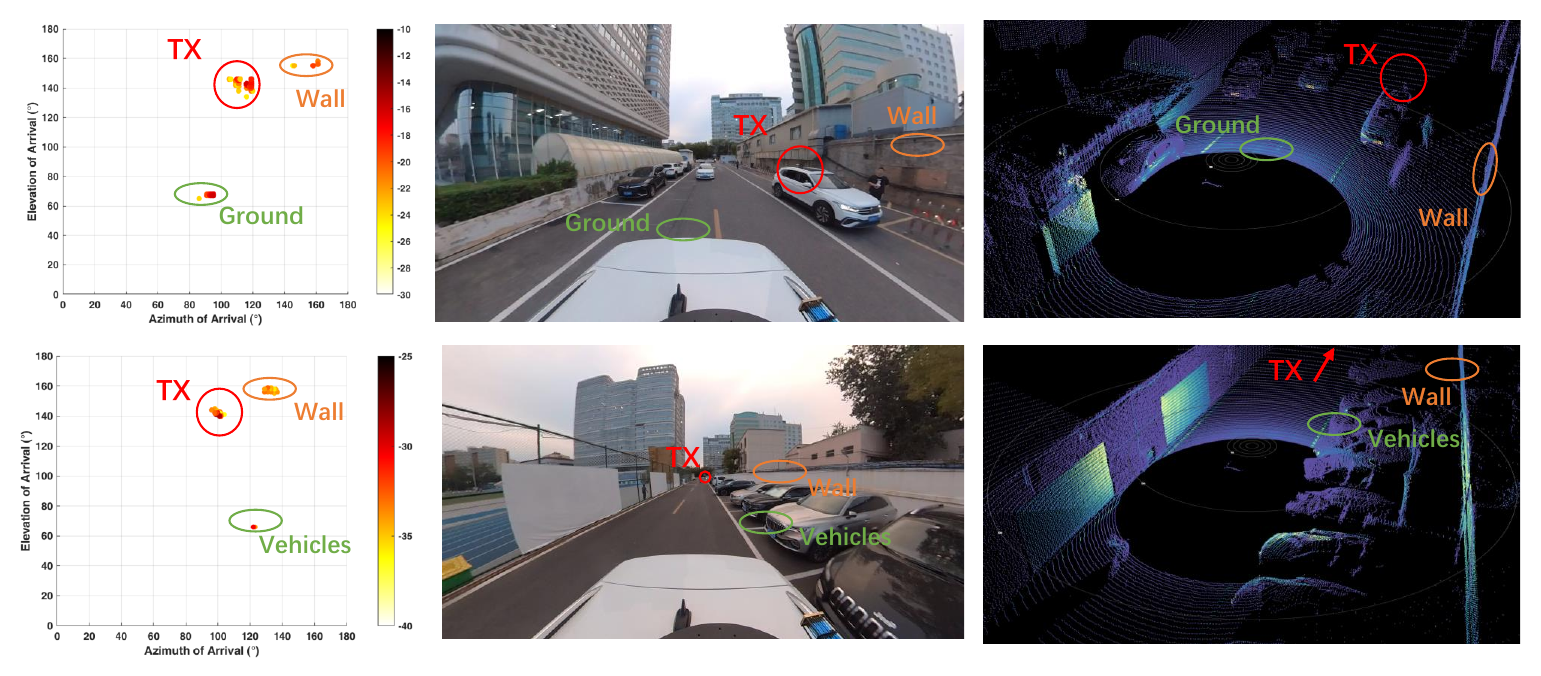} 
	\caption{Multi-modal verification combining wireless channel sounding, camera, and LiDAR data.}
	\label{Multidata}
\end{figure*}

Power delay profile (PDP) is extensively employed to characterize the power levels of received paths with propagation delays and to describe the distribution of multipath components in measured environments.
The instantaneous PDP  $P(t,\tau )$ is denoted as
\begin{equation} 
\label{math4}
P(t,\tau ) = {\left| {h(t,\tau )} \right|^2}.
\end{equation}

Fig. \ref{multipath}(a) illustrates the PDP evolution over the measurement campaign, where the horizontal axis denotes delay and the vertical axis indicates snapshot index. A dominant LoS component can be clearly observed, continuously tracked across more than 800 snapshots. In addition, several weaker NLoS components appear intermittently at different times, resulting in significant delay spreads. Fig. \ref{multipath}(b) shows a representative snapshot at 40 s, where 7 clusters are identified. The earliest and strongest clusters (Cluster 1 and 2) correspond to the LoS component and early reflections, while the later clusters (e.g., Cluster 5–7) represent weak reflections with delays exceeding 200 ns. This clearly demonstrates the extremely high temporal resolution, enabling the identification of multipath components at the nanosecond scale.

\subsection{Angle Estimation Verification}

The angular resolution capability of the developed platform is further validated by applying the Space-Alternating Generalized Expectation-maximization (SAGE) algorithm to extract the angles of arrival (AoAs). 
Fig. \ref{multipath}(c) shows the estimated azimuth over the measurement duration, with color indicating the relative power of multipath components. The azimuthal distribution is primarily concentrated within $80^{\circ}$--$140^{\circ}$, with a dominant and stable LoS trajectory observed near $100^{\circ}$ throughout the $30$--$100 \ \text{s}$ measurement period. In addition to this strong component, several weaker reflections appear intermittently around $70^{\circ}$ and $150^{\circ}$, reflecting the influence of dynamic roadside scatterers. Notably, even when weaker paths exhibit a power difference exceeding $20$--$30 \ \text{dB}$ compared with the dominant LoS, the system still resolves them. This confirms the  high dynamic range and angular resolution.

\subsection{Multi-Modal Data Verification}

To further validate the reliability and interpretability of the developed platform, we have conducted multi-modal data fusion by integrating wireless measurement results with synchronized camera and LiDAR information. 
Fig. \ref{Multidata} presents two representative snapshots at $5 \ \text{s}$ and $20 \ \text{s}$. At 5  s, the LoS component is clearly observed, with an elevation angle of approximately $150^{\circ}$, corresponding to  TX. 
In addition, a prominent reflection cluster arises from the building wall on the right-hand side, while the ground-reflected components are concentrated within the elevation range of $40^{\circ}$--$60^{\circ}$. 
The camera and LiDAR evidence enable a geometry-consistent association between scattering objects and  measurement results. At $20 \ \text{s}$, the LoS component is still present, but compared with the $5 \ \text{s}$ snapshot, its angular position shifts slightly toward the center. This change is consistent with the relative motion of Rx  with respect to Tx, demonstrating the  capability of dynamic LoS tracking. Meanwhile, beyond the wall reflection, additional multipath clusters corresponding to parked vehicles appear in the $60^{\circ}$--$80^{\circ}$ elevation region, with power levels about $20 \ \text{dB}$ weaker than the LoS but still reliably detected. 

It should be noted that the objective of the proposed framework is to establish a physically interpretable association between dominant channel clusters and environmental features using synchronized sensing data, rather than provide ground-truth identification of individual multipath components. The association is verified through geometric consistency between the measured cluster parameters (delay and AoA) and the propagation geometry reconstructed from synchronized LiDAR point clouds and panoramic images. Although rigorous accuracy evaluation would require an independent ground-truth reference, the association uncertainty is bounded by the sensing and channel sounding errors, including a LiDAR ranging accuracy of ±5 cm, a delay estimation uncertainty below 0.1 ns, and a residual AoA estimation error below  $1^\circ$. 
Consequently, the resulting uncertainty is sufficiently small compared with the characteristic dimensions of dominant environmental scatterers, enabling reliable physical interpretation of major channel clusters.

\subsection{Discussion}

The above results demonstrate that the proposed system provides an integrated, scalable, and environment-aware measurement capability beyond standalone channel sounding or sensing systems. 
Compared with conventional channel sounders, the proposed system supports calibrated RF channel measurements over 10 MHz--44 GHz, with 1 ns  delay resolution and an approximately  130 dB  measurable propagation loss at  28 GHz. 
Its modular NI PXIe-based architecture and reconfigurable RF front end further enable rapid adaptation to different frequency bands and measurement scenarios by replacing key RF modules, such as up/down converters, power amplifiers, switches, and antenna arrays.

Compared with standalone sensing platforms, the proposed system additionally provides radio-channel observability, including PL, PDP, AoA, and dominant channel clusters. Meanwhile, synchronized LiDAR point clouds, panoramic images, GNSS data, and IMU-assisted motion information provide environmental awareness for physical interpretation of the measured channels. The LiDAR subsystem achieves a ranging accuracy of approximately $\pm5 $ cm, and the destaggering correction reduces the angular error of vertical structures from $1.2^\circ\pm0.3^\circ$ to $0.08^\circ\pm0.02^\circ$. In typical urban V2I scenarios with vehicle speeds of 40–60 km/h, a  50 ms  frame-level alignment mismatch corresponds to only  0.56–0.83 m  spatial displacement. 
These metrics demonstrate that the proposed platform can provide both accurate channel measurements and reliable environmental context, enabling cluster-level association between measured propagation characteristics and physical environmental features.

Therefore, the measured  results jointly validate the platform in terms of large-scale attenuation characterization, delay- and angular-domain multipath resolution, and environment-aware interpretation. These capabilities are highly relevant to 6G vehicular communications, where high-frequency propagation, mobility-induced channel dynamics, and sensing-aided channel modeling are expected to be fundamental. 
Consequently, the proposed platform provides a practical experimental basis for 6G-oriented environment-aware channel modeling, ISAC research, and AI-assisted channel prediction.

\section{CONCLUSION}
In this paper, we present a multi-modal environment sensing and channel sounding system designed to address the limitations of traditional channel sounders in 6G dynamic scenarios. 
By integrating diverse sensing devices and channel sounders, the system enables temporally and spatially synchronized acquisition of multi-modal data, including images, point clouds, GNSS data, and multi-band channel data.  
The system supports  10 MHz--44 GHz  operation with up to  1 GHz  bandwidth and  1 ns  delay resolution, while its modular RF front end allows rapid reconfiguration for different frequency bands and measurement scenarios.
The metrological performance of the platform is verified through calibration, system characterization, and a representative  28 GHz  V2I measurement campaign. The results demonstrate an approximately  130 dB  measurable propagation loss at  28 GHz, stable phase-coherent acquisition, calibrated channel response, and reliable LiDAR-based environmental reconstruction with $\pm5\,\mathrm{cm}$ ranging accuracy. 
The synchronized multi-modal data further enable physically interpretable association between dominant channel clusters and large-scale environmental features. 
These capabilities provide a practical experimental foundation for 6G-oriented environment-aware channel modeling, ISAC research, and AI-assisted channel prediction.

   \balance
   \bibliographystyle{IEEEtran}

   \nocite{*}
   
   \bibliography{IEEEabrv,ref}

\begin{thebibliography}{10}
\providecommand{\url}[1]{#1}
\csname url@samestyle\endcsname
\providecommand{\newblock}{\relax}
\providecommand{\bibinfo}[2]{#2}
\providecommand{\BIBentrySTDinterwordspacing}{\spaceskip=0pt\relax}
\providecommand{\BIBentryALTinterwordstretchfactor}{4}
\providecommand{\BIBentryALTinterwordspacing}{\spaceskip=\fontdimen2\font plus
\BIBentryALTinterwordstretchfactor\fontdimen3\font minus
  \fontdimen4\font\relax}
\providecommand{\BIBforeignlanguage}[2]{{%
\expandafter\ifx\csname l@#1\endcsname\relax
\typeout{** WARNING: IEEEtran.bst: No hyphenation pattern has been}%
\typeout{** loaded for the language `#1'. Using the pattern for}%
\typeout{** the default language instead.}%
\else
\language=\csname l@#1\endcsname
\fi
#2}}
\providecommand{\BIBdecl}{\relax}
\BIBdecl

\bibitem{he2024}
R.~He and B.~Ai, \emph{Wireless Channel Measurement and Modeling in Mobile
  Communication Scenario: Theory and Application}.\hskip 1em plus 0.5em minus
  0.4em\relax CRC Press, 2024.

\bibitem{zzy2023}
Z.~Zhang \emph{et~al.}, ``A general channel model for integrated sensing and
  communication scenarios,'' \emph{IEEE Commun. Mag.}, vol.~61, no.~5, pp.
  68--74, 2023.

\bibitem{lei2023}
L.~Zhang, X.~Chu, and M.~Zhai, ``Machine learning-based integrated wireless
  sensing and positioning for cellular network,'' \emph{IEEE Trans. Instrum.
  Meas.}, vol.~72, pp. 1--11, 2023.

\bibitem{itu1}
\BIBentryALTinterwordspacing
{ITU-R\vspace{0mm}}, ``{Future technology trends of terrestrial International
  Mobile Telecommunications systems towards 2030 and beyond},'' \emph{{Report
  M.2516-0}}, Nov. 2022. [Online]. Available:
  \url{https://www.itu.int/dms_pub/itu-r/opb/rep/R-REP-M.2516-2022-PDF-E.pdf}
\BIBentrySTDinterwordspacing

\bibitem{hbx}
B.~He \emph{et~al.}, ``Toward secure semantic transmission in the era of
  {GenAI: A} diffusion-based framework,'' \emph{IEEE Commun. Mag.}, vol.~64,
  no.~7, pp. 110--117, 2026.

\bibitem{COST}
\BIBentryALTinterwordspacing
{COST CA20120 Action}, ``{INTERACT}: {Intelligence}-enabling radio
  communications for seamless inclusive interactions,'' 2021. [Online].
  Available: \url{https://www.cost.eu/actions/CA20120}
\BIBentrySTDinterwordspacing

\bibitem{zxj2025-2}
X.~Zhang, R.~He, M.~Yang, Z.~Zhang, Z.~Qi, and B.~Ai, ``Vision aided channel
  prediction for vehicular communications: {A} case study of received power
  prediction using {RGB} images,'' \emph{IEEE Trans. Veh. Technol.}, vol.~74,
  no.~11, pp. 17\,531--17\,544, 2025.

\bibitem{mao2024}
K.~Mao \emph{et~al.}, ``A survey on channel sounding technologies and
  measurements for {UAV}-assisted communications,'' \emph{IEEE Trans. Instrum.
  Meas.}, vol.~73, pp. 1--24, 2024.

\bibitem{zxj2025}
X.~Zhang, R.~He, M.~Yang, Z.~Qi, Z.~Zhang, B.~Ai, and Z.~Zhong, ``Vision-aided
  channel prediction based on image segmentation at street intersection
  scenarios,'' \emph{IEEE Trans. Cogn. Commun. Netw.}, vol.~12, pp. 1678--1693,
  2026.

\bibitem{gu2026}
J.~Gu, L.~Yin, X.~Zhang, and H.~Yang, ``{SDR}-based bidirectional precise
  wireless channel sounding for physical layer key generation,'' \emph{IEEE
  Trans. Instrum. Meas.}, vol.~75, pp. 1--20, 2026.

\bibitem{imran2024}
S.~Imran, G.~Charan, and A.~Alkhateeb, ``Environment semantic communication:
  {Enabling} distributed sensing aided networks,'' \emph{IEEE Open J. Commun.
  Soc.}, vol.~5, pp. 7767--7786, 2024.

\bibitem{feifei2023}
Y.~Yang, F.~Gao, X.~Tao, G.~Liu, and C.~Pan, ``Environment semantics aided
  wireless communications: {A} case study of {mmWave} beam prediction and
  blockage prediction,'' \emph{{IEEE} J. Sel. Areas Commun.}, vol.~41, no.~7,
  pp. 2025--2040, 2023.

\bibitem{wang2026}
M.~Wang, H.~Wang, H.~Dong, and A.~E. Saddik, ``{3D Wi-Fi} signal measurement in
  realistic digital twin testbed environments using ray tracing,'' \emph{IEEE
  Trans. Instrum. Meas.}, vol.~75, pp. 1--13, 2026.

\bibitem{bl2025}
L.~Bai, Z.~Huang, M.~Sun, X.~Cheng, and L.~Cui, ``Multi-modal intelligent
  channel modeling: {A} new modeling paradigm via synesthesia of machines,''
  \emph{IEEE Commun. Surveys Tuts.}, vol.~28, pp. 2612--2649, 2026.

\bibitem{cx-2024}
X.~Cheng \emph{et~al.}, ``Intelligent multi-modal sensing-communication
  integration: {Synesthesia} of machines,'' \emph{IEEE Commun. Surveys Tuts.},
  vol.~26, no.~1, pp. 258--301, 2024.

\bibitem{deepsense}
A.~Alkhateeb, G.~Charan, T.~Osman, A.~Hredzak, J.~Morais, U.~Demirhan, and
  N.~Srinivas, ``{DeepSense 6G}: A large-scale real-world multi-modal sensing
  and communication dataset,'' \emph{{IEEE} Commun. Mag.}, vol.~61, no.~9, pp.
  122--128, 2023.

\bibitem{CJE}
Z.~Zhang \emph{et~al.}, ``Non-stationarity characteristics in dynamic vehicular
  {ISAC} channels at 28 {GHz},'' \emph{Chinese J. Electron.}, vol.~34, no.~1,
  pp. 73--81, 2025.

\bibitem{sm2025}
M.~Sandra, C.~Nelson, X.~Li, X.~Cai, F.~Tufvesson, and A.~J. Johansson, ``A
  wideband distributed massive {MIMO} channel sounder for communication and
  sensing,'' \emph{IEEE Trans. Antennas Propag.}, vol.~73, no.~4, pp.
  2074--2085, 2025.

\bibitem{wang2023usrp}
Y.~Wang, W.~Wang, Y.~Wu, J.~Liu, Q.~Zhang, J.~Wang, and W.~Fan, ``{USRP}-based
  multifrequency multiscenario channel measurements and modeling for {5G}
  campus internet of things,'' \emph{IEEE Internet Things J.}, vol.~11, no.~8,
  pp. 13\,865--13\,883, 2023.

\bibitem{bas2018real}
C.~U. Bas, V.~Kristem, R.~Wang, and A.~F. Molisch, ``Real-time ultra-wideband
  channel sounder design for 3--18 {GHz},'' \emph{IEEE Trans. Commun.},
  vol.~67, no.~4, pp. 2995--3008, 2018.

\bibitem{cxs2024}
X.~Cai, E.~L. Bengtsson, O.~Edfors, and F.~Tufvesson, ``A switched array
  sounder for dynamic millimeter-wave channel characterization: {Design},
  implementation, and measurements,'' \emph{IEEE Trans. Antennas Propag.},
  vol.~72, no.~7, pp. 5985--5999, 2024.

\bibitem{bas2019}
C.~U. Bas \emph{et~al.}, ``Real-time millimeter-wave {MIMO} channel sounder for
  dynamic directional measurements,'' \emph{IEEE Trans. Veh. Technol.},
  vol.~68, no.~9, pp. 8775--8789, 2019.

\bibitem{cai2020dynamic}
X.~Cai, G.~Zhang, C.~Zhang, W.~Fan, J.~Li, and G.~F. Pedersen, ``Dynamic
  channel modeling for indoor millimeter-wave propagation channels based on
  measurements,'' \emph{IEEE Trans. Commun.}, vol.~68, no.~9, pp. 5878--5891,
  2020.

\bibitem{al2024}
A.~Al-Ameri, J.~Sanchez, F.~Tufvesson, and X.~Cai, ``A fast rotating-mirror
  sounder for dynamic millimeter-wave channel characterization,'' in
  \emph{Proc. IEEE 100th Veh. Technol. Conf.}, 2024, pp. 1--5.

\bibitem{huang2020multi}
J.~Huang, C.-X. Wang, H.~Chang, J.~Sun, and X.~Gao, ``Multi-frequency
  multi-scenario millimeter wave {MIMO} channel measurements and modeling for
  {B5G} wireless communication systems,'' \emph{IEEE J. Sel. Areas Commun.},
  vol.~38, no.~9, pp. 2010--2025, 2020.

\bibitem{mao2023uav}
K.~Mao \emph{et~al.}, ``A {UAV}-aided real-time channel sounder for highly
  dynamic nonstationary {A2G} scenarios,'' \emph{IEEE Trans. Instrum. Meas.},
  vol.~72, pp. 1--15, 2023.

\bibitem{chen2023passive}
C.~Chen, D.~Fei, P.~Zheng, and B.~Ai, ``A passive channel measurement and
  analysis based on a {5G} commercial network in {V2I} communications,''
  \emph{Electronics}, vol.~12, no.~17, p. 3715, 2023.

\bibitem{wu2021measurement}
T.~Wu, X.~Yin, L.~Zhang, and J.~Ning, ``Measurement-based channel
  characterization for {5G} downlink based on passive sounding in sub-{6 GHz
  5G} commercial networks,'' \emph{IEEE Trans. Wireless Commun.}, vol.~20,
  no.~5, pp. 3225--3239, 2021.

\bibitem{miao2023sub}
H.~Miao, J.~Zhang, P.~Tang, L.~Tian, X.~Zhao, B.~Guo, and G.~Liu, ``Sub-6 {GHz}
  to {mmWave} for {5G-advanced} and beyond: {Channel} measurements,
  characteristics and impact on system performance,'' \emph{IEEE J. Sel. Areas
  Commun.}, vol.~41, no.~6, pp. 1945--1960, 2023.

\bibitem{kim2022}
M.~Kim, H.~Tsukada, K.~Kumakura, R.~Takahashi, N.~Suzuki, H.~Sawada, and
  T.~Matsumura, ``A {24/60-GHz} dual-band double-directional channel sounder
  using {COTS} phased arrays,'' in \emph{Proc. IEEE Int. Conf. Commun.
  Workshops (ICC Workshops)}, 2022, pp. 1113--1117.

\bibitem{gen2024}
C.~Gentile, J.~Senic, A.~Bodi, S.~Berweger, R.~Caromi, and N.~Golmie,
  ``Context-aware channel sounder for {AI}-assisted radio-frequency channel
  modeling,'' in \emph{Proc. 18th Eur. Conf. Antennas Propag. (EuCAP)}, 2024,
  pp. 1--5.

\bibitem{Charan2024}
G.~Charan and A.~Alkhateeb, ``User identification: {A} key enabler for
  multi-user vision-aided communications,'' \emph{IEEE Open J. Commun. Soc.},
  vol.~5, pp. 472--488, 2024.

\bibitem{fd2022}
D.~Fei, C.~Chen, P.~Zheng, D.~Zhang, J.~Yang, H.~Chen, and B.~Ai, ``A novel
  millimeter-wave channel measurement platform for {6G} intelligent railway
  scenarios,'' \emph{China Commun.}, vol.~19, no.~11, pp. 60--73, 2022.

\bibitem{gong2025}
H.~Gong, X.~Wang, C.~Liu, Y.~Zhao, Y.~Liu, N.~Li, and W.~Yang, ``Fusion of
  measurement and simulation technique for electromagnetic environment analysis
  in large-scale urban areas,'' \emph{IEEE Trans. Instrum. Meas.}, vol.~74, pp.
  1--12, 2025.

\bibitem{QI2023241}
Z.~Qi \emph{et~al.}, ``Point cloud-based environment reconstruction and ray
  tracing simulations for railway tunnel channels,'' \emph{High-speed Railway},
  vol.~1, no.~4, pp. 241--247, 2023.

\bibitem{Nishio2021}
T.~Nishio, Y.~Koda, J.~Park, M.~Bennis, and K.~Doppler, ``When wireless
  communications meet computer vision in beyond {5G},'' \emph{IEEE Commun.
  Standards Mag.}, vol.~5, no.~2, pp. 76--83, 2021.

\bibitem{bae2025}
K.-B. Bae, Y.~Yoo, S.~Park, and S.-O. Park, ``A novel channel sounding method
  based on pulsed {LFM} signal for {mm-Wave} application: {Instrumentation},
  measurement, and analysis,'' \emph{IEEE Trans. Instrum. Meas.}, vol.~74, pp.
  1--16, 2025.

\end{thebibliography}

 \begin{IEEEbiography}[{\includegraphics[width=1in,height=1.25in,clip,keepaspectratio]{XuejianZhang.pdf}}]{Xuejian Zhang}(Student Member, IEEE)
	received the B.S. degree in communication engineering from Lanzhou Jiaotong University, Lanzhou, China, in 2021. 
	He is currently pursuing the Ph.D. degree with the School of Electronics and Information Engineering, Beijing Jiaotong University, Beijing, China. 
	From August to December 2025, he was a Visiting Scholar with the Institute of Science and Technology, Niigata University,   Japan.
	His research interests include wireless channel measurement and modeling, railway and vehicular communications, and vision-aided intelligent channel prediction.
\end{IEEEbiography}

\vspace{-1.35   cm}

\begin{IEEEbiography}[{\includegraphics[width=1in,height=1.25in,clip,keepaspectratio]{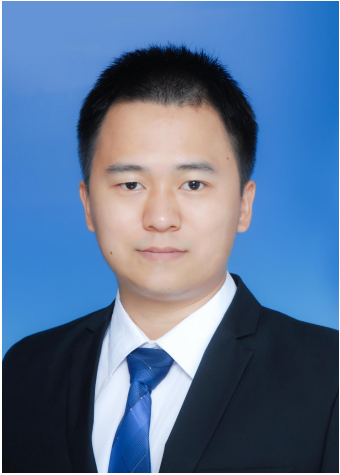}}]{Ruisi He}(Senior Member, IEEE) received the B.E. and Ph.D. degrees from Beijing Jiaotong University (BJTU), Beijing, China, in 2009 and 2015, respectively.
Dr. He is currently a Professor with the School of Electronics and Information Engineering, BJTU. 
Dr. He has been a Visiting Scholar in Georgia Institute of Technology, USA, University of Southern California, USA, and Universit\'e Catholique de Louvain, Belgium. His research interests include wireless propagation channels, 5G and 6G communications. He has authored/co-authored 8 books, 5 book chapters, more than 200 journal and conference papers, as well as several patents.

Dr. He has been an Editor of the \emph{IEEE Transactions on Communications}, the \emph{IEEE Transactions on Wireless Communications}, the \emph{IEEE Transactions on Antennas and Propagation}, the \emph{IEEE Antennas and Propagation Magazine}, the \emph{IEEE Communications Letters}, the \emph{IEEE Open Journal of Vehicular Technology}, and a Lead Guest Editor of the \emph{IEEE Journal on Selected Areas in Communications} and the \emph{IEEE Transactions on Antennas and Propagation}. He served as the Early Career Representative (ECR) of Commission C, International Union of Radio Science (URSI). He received the URSI Issac Koga Gold Medal in 2021, the IEEE ComSoc Asia-Pacific Outstanding Young Researcher Award in 2019, the URSI Young Scientist Award in 2015, and several Best Paper Awards in IEEE journals and conferences.
\end{IEEEbiography}

\vspace{-1.4 cm}

\begin{IEEEbiography}[{\includegraphics[width=1in,height=1.25in,clip,keepaspectratio]{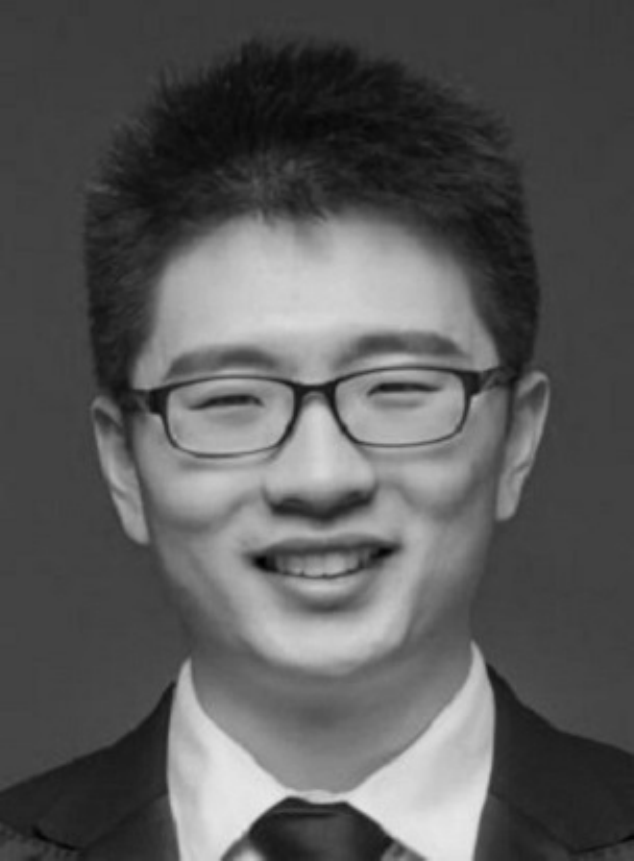}}]{Mi Yang}(Member, IEEE)
	received the M.S. and Ph.D. degrees from Beijing Jiaotong University, Beijing, China, in 2017 and 2021, respectively. He is currently an associate professor with the School of Electronics and Information Engineering, Beijing Jiaotong University. His research interests are focused on wireless channel measurement and modeling, vehicular and railway communications, and AI in channel research.
\end{IEEEbiography}

\vspace{-1.4 cm}

\begin{IEEEbiography}[{\includegraphics[width=1in,height=1.25in,clip,keepaspectratio]{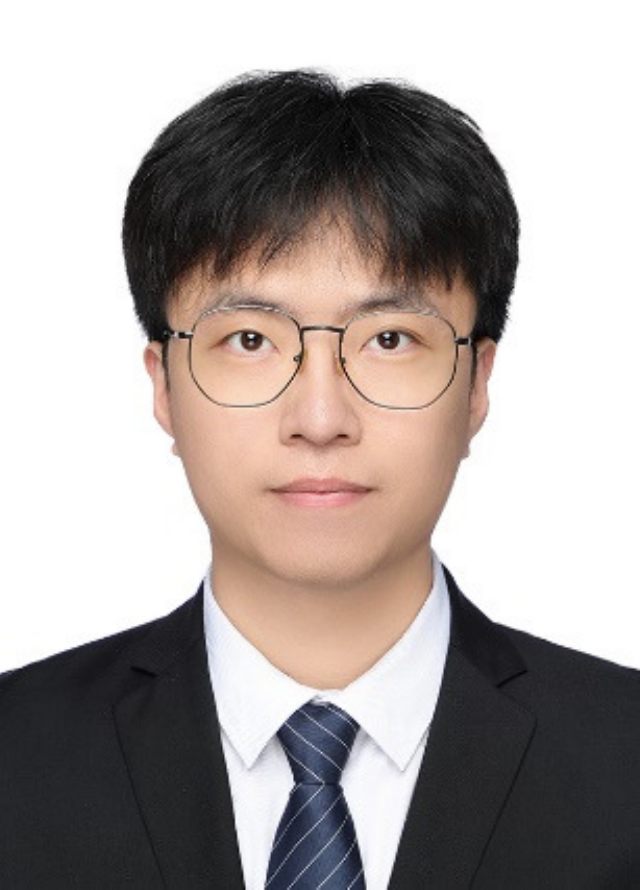}}]{Zhengyu Zhang}(Student Member, IEEE) 
received the B.S. and M.S. degrees in electronic and communication engineering and the Ph.D. degree in information and communication systems from Beijing Jiaotong University, China, in 2018, 2021, and 2025, respectively. From 2024 to 2025, he was a Visiting Ph.D. Student with the University of Bologna, Italy. He is currently a Lecturer with the School of Electronics and Information Engineering, Beijing Jiaotong University. His research
interests include ISAC, channel semantics, deep learning, and 6G wireless channel measurement and modeling.
\end{IEEEbiography}
\vspace{-3 cm}

\begin{IEEEbiography}[{\includegraphics[width=1in,height=1.25in,clip,keepaspectratio]{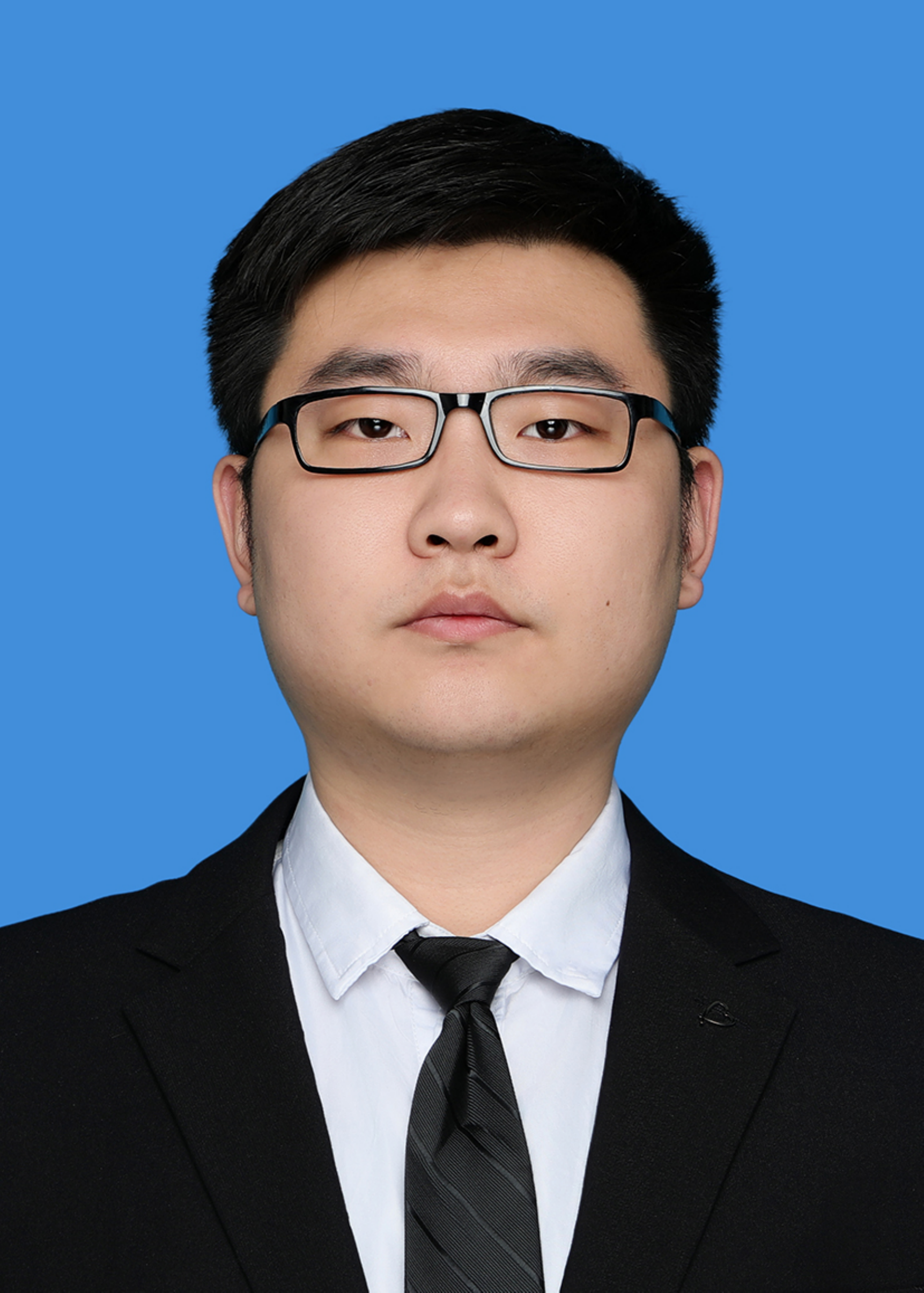}}]{Ziyi Qi}
	is currently pursuing the Ph.D. degree with the School of Electronics and Information Engineering, Beijing Jiaotong University, Beijing, China. His research interests include wireless channel measurement and modeling, wireless propagation environment reconstruction and deterministic channel modeling.
\end{IEEEbiography}

\end{document}